\documentclass{article}

\usepackage{arxiv}

\usepackage[utf8]{inputenc} % allow utf-8 input
\usepackage[T1]{fontenc}    % use 8-bit T1 fonts
\usepackage{hyperref}       % hyperlinks
\usepackage{url}            % simple URL typesetting
\usepackage{booktabs}       % professional-quality tables
\usepackage{amsfonts}       % blackboard math symbols
\usepackage{nicefrac}       % compact symbols for 1/2, etc.
\usepackage{microtype}      % microtypography
\usepackage{graphicx}
\usepackage{natbib}
\usepackage{doi}

\usepackage{multirow}%
\usepackage{amsmath,amssymb,amsfonts}%
\usepackage{amsthm}%
\usepackage{booktabs}%
\usepackage{siunitx}
\usepackage[version=3]{mhchem}
\DeclareSIUnit\angstrom{\text{\AA}}

\title{Explicit pore pre-conditioning and inverse design of amorphous nanoporous carbon}

\date{Novermber 1, 2026}	% Here you can change the date presented in the paper title
\author{ \href{https://orcid.org/0000-0002-3716-9163}{\includegraphics[scale=0.06]{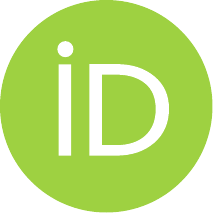}\hspace{1mm}Matteo~Alberghini}\thanks{Corresponding author. Mail-to: \texttt{matteo.alberghini@gemmate-technologies.com}} \\
	Gemmate Technologies s.r.l.\\
    Via Reano 31, 10090\\
	Buttigliera Alta (Turin), Italy \\
	\And
    \href{https://orcid.org/0000-0003-1973-0612}{\includegraphics[scale=0.06]{orcid.pdf}\hspace{1mm}Giacomo~Pierotti} \\	
    Gemmate Technologies s.r.l.\\
    Via Reano 31, 10090\\
	Buttigliera Alta (Turin), Italy \\
    \And
	\href{https://orcid.org/0000-0002-5093-0105}{\includegraphics[scale=0.06]{orcid.pdf}\hspace{1mm}Maryam~Amini} \\
	University of Turin\\
	Department of Chemistry\\
    Via Pietro Giuria 7, 10125\\
	Turin, Italy \\
    \And
	\href{https://orcid.org/0000-0002-0914-4217}{\includegraphics[scale=0.06]{orcid.pdf}\hspace{1mm}Mauro~Sgroi} \\
	University of Turin\\
	Department of Chemistry\\
    Via Pietro Giuria 7, 10125\\
	Turin, Italy \\
    \And
    \href{https://orcid.org/0000-0001-5220-2227}{\includegraphics[scale=0.06]{orcid.pdf}\hspace{1mm}Alessio~Tommasi} \\	
    Gemmate Technologies s.r.l.\\
    Via Reano 31, 10090\\
	Buttigliera Alta (Turin), Italy \\
}

\renewcommand{\headeright}{Inverse design of nanoporous carbon}
\renewcommand{\undertitle}{}
\renewcommand{\shorttitle}{}

\hypersetup{
pdftitle={Inverse design of nanoporous carbon},
pdfsubject={cond-mat.mtrl-sci},
pdfauthor={Matteo Alberghini, Giacomo Pierotti},
pdfkeywords={nanoporous carbon, inverse design, particle swarm optimization, molecular dynamics, identifiability},
}

\begin{document}
\maketitle

\begin{abstract}
	Porous carbons store energy, capture gases and purify water, and their performance depends strongly on the geometry of their pores. Atomistic models mostly simulate their synthesis by molecular dynamics, with results that depend on the interatomic potential, and return the pore network as an output rather than as a specification. Here we show that the pore network can be an input: pores drawn from a target size distribution fill a prescribed volume fraction, carbon is grown around them by stochastic ring placement, and a particle swarm recovers the construction parameters from a structural fingerprint. The search reproduces the density, surface area and pore volume of a carbon produced by quenched molecular dynamics within about 5 \%, together with its pore size distribution. As a stress test, it also recovers a bimodal distribution, and the search history reveals which parameters the targets constrain and flags a failed search. A short anneal completes the network, recovering the coordination of the reference while the pore size distribution keeps its shape. Molecular dynamics still supplies the local chemistry, but the pore geometry becomes a design variable for a given application.
\end{abstract}

% keywords can be removed
\keywords{nanoporous carbon \and inverse design \and pore size distribution \and particle swarm optimization \and molecular dynamics \and identifiability}

\section{Introduction}

Disordered nanoporous carbons are used in supercapacitor electrodes, gas storage and separation media, and water purification systems, where their performance depends on the geometry of the void they enclose as well as on the carbon itself. In carbide-derived carbons, the synthesis temperature tunes the average pore width within the sub-nanometer range \cite{gogotsi2003nanoporous}, and the anomalous rise in capacitance observed as the pore width falls below the diameter of the solvated ion showed that such pores contribute to charge storage, so that the pore size distribution (PSD), and not the surface area alone, sets the capacitance \cite{chmiola2006anomalous}. Across a series of commercial carbons, the capacitance was later found to follow the structural disorder rather than the pore size \cite{liu2024structural}, so that the pore geometry and the local structure of the carbon both matter. Beyond charge storage, molecular dynamics of a concentrated aqueous electrolyte confined in carbon slit pores shows the ionic current to vary non-monotonically with the pore width over a few ångström \cite{alberghini2025pore}, a response which no single average width can represent. The design-relevant variable is therefore the distribution of pore sizes, and a model of these materials is useful to the extent that it reproduces a specified void geometry together with the local structure of a disordered carbon.

Two established families of models, reviewed by Corrente et al. \cite{corrente2025slit}, do not meet this requirement. Idealised geometric models replace the disordered carbon with a prescribed shape, most commonly a slit pore bounded by rigid graphene walls \cite{alberghini2025pore, lu2017structure}: the pore geometry is exact and adsorption calculations are computationally inexpensive, but the skeleton carries none of the curvature, non-hexagonal rings or connectivity of a disordered carbon. The omission matters: inside atomistic carbide-derived carbon electrodes, Merlet et al. \cite{merlet2012molecular} obtained capacitances of 87 and 125 F g\textsuperscript{-1} for an ionic liquid, in agreement with experiment, whereas simulations of simpler electrode geometries had returned 6 to 40 F g\textsuperscript{-1}. Their two carbons, moreover, had nearly the same PSD but differed by 43 \% in capacitance, a difference they attributed to the local structure of the carbon. Reconstruction methods instead invert measured data: reverse Monte Carlo and its hybrid variant adjust atomic positions until the computed structure factor matches the measured one \cite{mcgreevy2001reverse, pikunic2003structural,nguyen2008new, farmahini2013structural}, and molecular dynamics with machine-learned potentials can now be biased toward diffraction and spectroscopic data directly \cite{zarrouk2026molecular}. The resulting structures agree with the data by construction, but they require measurements of a sample which must already exist, and the inversion is non-unique: a one-dimensional pair correlation underdetermines a three-dimensional network, and hybrid reverse Monte Carlo models of one carbon, fitted to the same diffraction data from different starting configurations, quench schemes and constraints, yield distinguishable structures \cite{farmahini2015hybrid}. The data these methods fit, moreover, constrain the local order of the carbon more than its pore network, whose PSD is compared with experiment after the reconstruction rather than imposed on it \cite{farmahini2015hybrid}.

The dominant route is mimetic molecular dynamics (MD), which produces a disordered carbon by simulating a physical process, in one of two protocols: quenched MD (QMD) melts a random configuration and cools it at a prescribed rate \cite{palmer2010modeling, thompson2017atomistic}, while annealing MD holds a configuration at high temperature long enough for the network to reorganise, at a cost of nanoseconds of trajectory per structure on systems of tens of thousands of atoms \cite{de2017structural, de2019transferability}. Machine-learned potentials have carried the same route to systems of over 10\textsuperscript{5} atoms \cite{wang2022structure} and to doped carbons \cite{liu2025molecular}. These methods produce physically relaxed structures, although their local order depends on the interatomic potential, since no two of the common carbon potentials give the same structure under the same protocol \cite{de2019transferability}. Their limitation is that the pore geometry is an output and not an input: the user specifies a density and a thermal schedule and receives the void structure that results. The density largely dictates the position and the width of the PSD peak \cite{wang2022structure}, and in annealing MD the structures depend on the annealing temperature and the density alone \cite{vallejos20233d}. The quench rate of QMD adds a further lever, slower quenches giving larger pores, but its effect saturates, the average pore size plateauing at $\sim$9 \AA{} below a threshold rate \cite{palmer2010modeling}. Matching a specified PSD by mimetic MD is therefore a search over a few protocol parameters, and no mimetic protocol has been shown to produce a shape outside the family they span, a bimodal distribution for instance.

Imposing the void geometry before or during the dynamics removes part of this constraint. Luo et al. \cite{luo2021virtual} and Zhou et al. \cite{zhou2025development} exclude carbon from virtual voids during a melt-quench, controlling the porosity and the mean pore size through the number and the radius of the voids, a single radius per model being chosen by hand, in the latter case until the surface area matches a measurement to within 2 \%. Waysenson et al. \cite{waysenson2026soft} replace the fixed voids with a mobile template inserted during a quench driven by a machine-learned potential, so that one characteristic length sets the pore size of each model while the structural disorder is kept constant across the series. Yu et al. \cite{yu2025machine} carve pores into a carbon quenched with a machine-learned potential, by superimposing sinusoidal waves of fixed wavelengths along random directions, spanning porosities of 10 to 90 \% and mean pore sizes of 5 to 60 \AA{}, the pore size being set independently of the density, but the resulting PSD which is controlled only through its mean and its variance. Peng et al. \cite{peng2026realistic} fit a random field of the same kind, with several characteristic lengths, to the small-angle X-ray scattering and the gas-adsorption porosity of a measured carbon, then match atoms to the resulting void map by reverse MD and an anneal, and the PSD measured by gas adsorption validates the model but is not imposed during its construction. Ugwumadu et al. \cite{ugwumadu2024self} come closest to an explicit construction, placing spherical voids before the carbon, but release them during the dynamics, so that the pore volume redistributes and the initial mesopores end as micropores. These works establish that porosity can be controlled constructively and that pore size can be decoupled from density; across all of them, however, the pore size enters as one or a few characteristic lengths, never as a target distribution, and it is reached by adjustment or by fitting another observable rather than by inverting the PSD itself.

%THIS WORK
Here we present a generator which takes as inputs the PSD and the volume fraction of the pores to be placed, together with an inverse-design procedure which recovers its parameters from a target structural fingerprint. Pores sampled from the target PSD, which may be a tabulated curve of arbitrary shape, are first placed as a sphere packing under periodic boundary conditions until their volume fraction is reached, and a carbon network is then grown ring by ring around them, each ring being accepted by a placement potential for the carbon-carbon interaction and by an atom-pore wall, which act as placement criteria rather than as a description of carbon energetics. Since the carbon changes the void, the porosity and the PSD of a structure are measured after its construction. The generation parameters are recovered by a particle swarm optimizer. The target may be a set of laboratory quantities or an atomistic reference, characterized through the same routines as the candidates, so that the two are compared under identical definitions. Construction, characterization and scoring are implemented in \verb+nanocarbon+, a Python library developed for this purpose, whose characterization routines are checked against independent tools for every channel but the pore connectivity, which has no external counterpart. The constructed structures are proxies, and a short anneal by MD, with the pores held as walls, completes their bond network while largely preserving the imposed void geometry.

We show that the search recovers the PSD of a QMD reference and, when the rings are placed with a harmonic bonded potential, the shape of a bimodal target which no other route has been shown to reach. We also analyze the search history, which identifies the generation parameters a target actually constrains and flags a search whose target was not reached. A short anneal, not optimized for this work, then brings the coordination, the skeletal density and the pore connectivity close to those of the reference, while the pore size distribution keeps its shape and shifts toward larger pores. The method is limited to elemental carbon, and a molecular dynamics stage remains part of the workflow. Within these limits, treating the pore geometry as a specification rather than as an outcome allows a carbon to be built for a given electrolyte or transport requirement rather than for a given synthesis route.

\section{Results and discussion}

\subsection{The pore network is set before the carbon is grown}\label{sec:method_overview}

\begin{figure}
    \centering
    \includegraphics[width=1\linewidth]{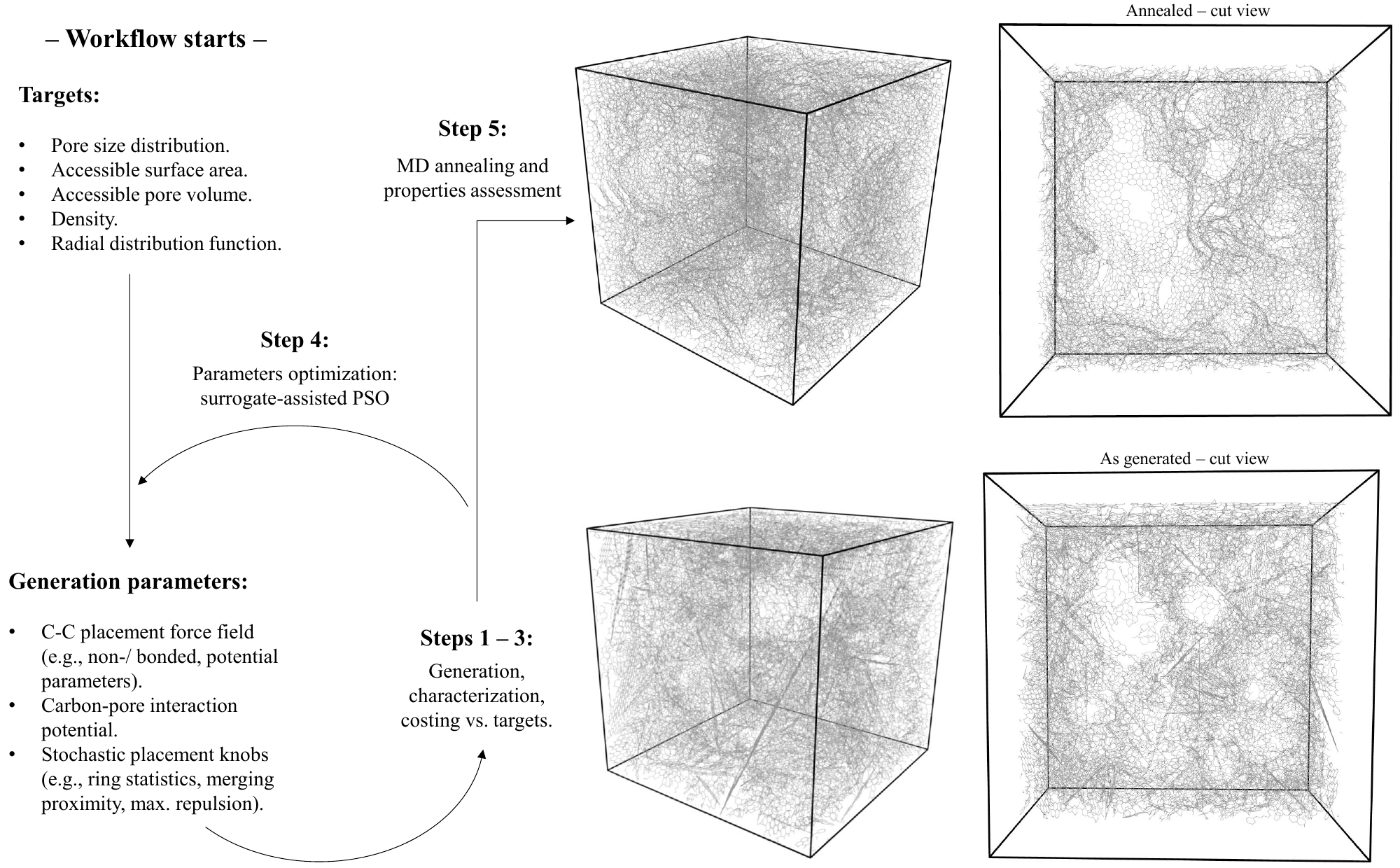}
    \caption{The inverse design of a nanoporous carbon structure. The targets are the structural properties the generated structure is required to reproduce, namely the pore size distribution, the accessible surface area, the accessible (probe-occupiable) pore volume, the density and the radial distribution function, covering global structure and local atomic order (all but the pore volume are scored in this work). The generation parameters are the knobs the construction exposes, namely the carbon-carbon placement potential, the carbon-pore wall and the stochastic placement rules. Steps 1 to 3 are the forward model: a proxy structure is built from one set of parameters, characterized and scored against the targets. Step 4 closes the loop, a particle swarm proposing the sets to evaluate next. Step 5 anneals the recovered optimal structure by molecular dynamics. The two cut views show the same structure as generated and after the anneal.}
    \label{fig:methods}
\end{figure}

In quenched molecular dynamics (QMD) the pore structure of an amorphous carbon is an outcome: a density and a thermal schedule are prescribed, and the pore size distribution (PSD) follows from them \cite{palmer2010modeling,thompson2017atomistic}. The methodology developed here inverts this relationship: the pores are placed from the target PSD and the carbon is grown around them to the target density, while a short relaxation at high temperature corrects the local bonding and leaves the pore network in place. The route consists of the five steps of Figure \ref{fig:methods}: construction (Step 1), characterization (Step 2), scoring (Step 3), search over the construction parameters (Step 4) and anneal by molecular dynamics (Step 5). The constructed structures are referred to as proxy structures, since they reproduce the targeted geometry without yet carrying a physical bond network, and as annealed structures after Step 5.
In Step 1 the void is placed before the carbon. Pore spheres with diameters drawn from the target PSD are inserted into the periodic cell until a pore volume fraction $\phi$ is reached, and carbon is then grown around them ring by ring until the target density $\rho$ is reached. A ring is accepted only if none of its atoms enters a pore and its energy stays below a threshold, this energy being the sum of a placement potential for the carbon-carbon interaction and of a Lennard-Jones wall for the carbon-pore one (see Section \ref{sec:structure_generation}). Two placement potentials are used in this work, a harmonic bonded model, which sets the geometry of a new ring with respect to the sheet it grows from, and the Tersoff bond-order potential \cite{tersoff1989modeling}. Under the bond exclusion used in this work, the Tersoff energy of a candidate ring leaves out its anchors, the two atoms of the network it is attached to, and every atom within 15 bonds of them, and so acts only on the contacts between the ring and distant parts of the network (see Section \ref{sec:method_potential}). This construction, referred to below as the Tersoff-contact construction, therefore carries no local bonding term: the local bonding is set only by the shape of the placed rings and by the acceptance test, while the pore wall and these contacts hold the carbon in place. Since the anneal of Step 5 replaces the placement potential entirely, its parameters need not take physical values, and together with those of the pore wall they become parameters of the construction which the search can tune.

Five of these parameters are searched here, selected on physical grounds, although the methodology leaves this choice free. The pore volume fraction $\phi$ sets how much void is placed. The depth $A$ and the range $B$ of the pore wall set how strongly and how far from each pore surface the carbon is held, the closest distance at which a ring is accepted being referred to below as the standoff. The dihedral stiffness $k_\varphi$ of the harmonic model sets how easily a new ring leaves the plane of its parent, and so how much curvature the network admits. The merge coefficient $m$ sets the distance, in units of the equilibrium bond length, below which a new atom is fused with an existing one, and so whether neighboring rings join into continuous sheets. The other constants take literature or working values (Supplementary Note 1). The pores placed, however, are not the pores measured: the wall keeps the carbon at a distance from every pore surface and the network leaves interstitial voids of its own, so that the measured porosity and PSD depart from the placed ones in a way which depends jointly on $\phi$, on the wall and on the placement potential. The map from the parameters to the measured structure has no analytical inverse, and Steps 2 to 4 explore it numerically.

In Step 2 each proxy structure is characterized with the same routines applied to the targets. The cost of the PSD routine grows linearly with the size of the structure, and it takes 9.4 min on a 100 \AA{} structure, against 2.7 h for a PoreBlazer run on the same grid (see Section \ref{sec:post-process}), so that every construction of a search can be characterized. In Step 3 the scored properties, among them the accessible surface area (ASA) and the radial distribution function (RDF), are combined into a single cost, the sum of their dimensionless deviations from the targets, which is zero for a structure reproducing every target (see Section \ref{sec:costing}). The RDF term is local, computed around points sampled at random in the cell, so that the least ordered regions are not masked by the well-formed ones. Two constructions from the same parameters differ in their random seed, and so do their costs, so that each candidate is built in seven replicates and scored by their mean cost.

Step 4 searches the parameters with a particle swarm optimizer (PSO) \cite{kennedy1995particle}, as already done to calibrate coarse-grained force fields against experimental data \cite{lombardo2020accelerated}. Each particle of the swarm is a set of parameters, which at every iteration moves under three pulls: its own inertia, the best position it has found and the best position found by the whole swarm (see Section \ref{sec:pso}). The swarm needs no gradient, and so tolerates a cost which is noisy and undefined wherever a construction fails. A quadratic model of the cost, refitted on the costs collected so far, proposes one further candidate per iteration. Every evaluation is retained, and Section \ref{sec:identifiability} reads from this history which parameters the targets constrain.

Finally, the recovered proxy structure is annealed (Step 5) with the molecular dynamics engine LAMMPS \cite{thompson2022lammps}, to cure the local topology which the construction leaves incomplete. The pores are kept as purely repulsive walls, cut from the same wall potential used in the construction, and the carbon is described by the environment-dependent interatomic potential (EDIP) \cite{marks2000generalizing} at 2500 K, both choices being replaceable without affecting the rest of the chain.

\subsection{The inverse design converges and its history identifies the constrained parameters}\label{sec:identifiability}

This section tests whether the search of Step 4 recovers its targets and what its history reveals about the searched parameters. Every search improves on its random start by more than the noise of the construction and then samples a set of statistically equivalent solutions. Their history shows that the targets constrain $\phi$ in three of the four runs and $B$ in one, whereas $k_\varphi$, $A$ and $m$ are left free, a reading which also flags the one run whose target was not reached.

The search was tested on a four campaigns combining two targets with two placement potentials. The first target is realistic: the properties of a 100 \AA{} structure produced by QMD (see Section \ref{sec:qmd_protocol}), characterized with the same routines as the proxy structures. Since amorphous carbon has no unique equilibrium structure, this reference is taken as a fully characterized target, against which the proxy structures can also be compared on the properties that were not scored, a comparison which a measured PSD would not allow. Its PSD has no sharp peak but a broad window, 80 \% of its pore volume lying between 4.5 and 11.9 \AA{}, which covers the sub-nanometer range where the anomalous capacitance of carbide-derived carbons is observed \cite{chmiola2006anomalous}. The second target is deliberately extreme: a synthetic bimodal PSD, with modes of equal volume at 7 and 16 \AA{} and the same density as the first, chosen as a synthetic stress test rather than taken from a specific experimental dataset. Its small mode makes it a stress test, for two reasons. At equal volume, the 7 \AA{} mode requires about twelve times as many pores as the 16 \AA{} one, $(16/7)^3$, the cube of the ratio of the two mode diameters. Moreover, the carbon network leaves interstitial voids of comparable size, which the measured PSD counts in the small mode together with the placed pores. A construction which recovers this target should also recover a unimodal one.

Each campaign is labeled by its target, Q for the quench-MD structure and S for the synthetic PSD, and by its placement potential, H for harmonic and T for Tersoff-contact, giving runs Q-H, Q-T, S-H and S-T. The Tersoff-contact runs, which have no dihedral term, search the other four parameters. All four campaigns were given the same budget: a random initial swarm of 16 particles followed by 25 iterations, i.e., 416 candidates, each built in seven replicates (see Section \ref{sec:costing}), close to 3000 constructions per campaign. The search ranges, chosen deliberately broad, are listed in Supplementary Note 1.

\subsubsection{The searches converge beyond the noise of the construction}\label{sec:convergence}
\begin{figure}
    \centering
    \includegraphics[width=1\linewidth]{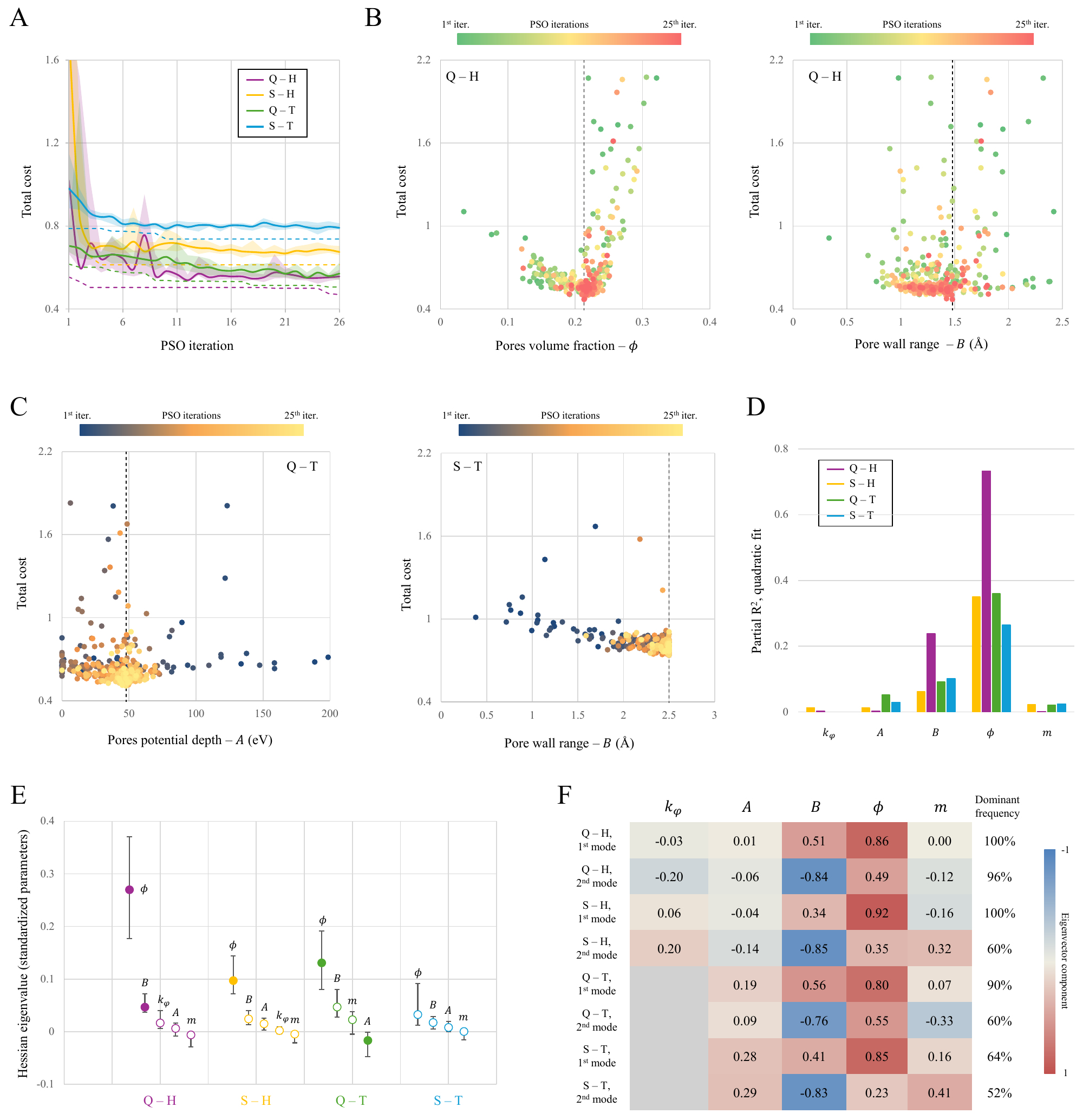}
    \caption{\textbf{Convergence and identifiability of the four inverse-design campaigns.} (A) Cost against iteration for runs Q-H, S-H, Q-T, and S-T; iteration 0 is the random initial swarm. Solid lines: median cost of the admissible candidates of the swarm. Shaded bands: interquartile range, i.e., the interval containing the central half of the swarm costs, a measure of how far the particles are from agreeing. Dashed lines: cost of the best candidate found so far. (B) Cost of candidates of run Q-H against the pore volume fraction $\phi$ and the pore-wall range $B$, colored by iteration; dashed lines mark the recovered values. (C) As (B), for the pore-wall depth $A$ of run Q-T and the pore-wall range $B$ of run S-T. (D) Share of the cost variance explained by each parameter alone (drop-one partial $R^2$, see Methods). (E) Eigenvalues of the Hessian of the fitted cost surface, i.e., its principal curvatures, ordered from the largest (the stiffest direction), with 95 \% intervals from a bootstrap over the particle trajectories, i.e., resampling whole particle histories (see Methods); a filled marker denotes a consistent attribution, the same parameter dominating the direction in at least 85 \% of the resamples. (F) Signed components of the eigenvectors of the two largest eigenvalues of each run, i.e., of its two stiffest directions, the sign chosen so that $\phi$ is positive, with the fraction of resamples in which the dominant parameter is unchanged.}
    \label{fig:results_pso}
\end{figure}
Figure \ref{fig:results_pso}A follows each search through two curves per run: the median cost of the swarm (solid lines), which records where the swarm as a whole sits, with the interquartile range of its costs as a shaded band, and the cost of the best candidate found so far (dashed lines). In every run the median falls fastest in the first iterations and then flattens, completing 90 \% of its total drop within 2 to 6 iterations in runs Q-H, S-H and S-T and within 15 iterations in run Q-T. Over the whole search the median falls by 45 and 61 \% of its initial value in runs Q-H and S-H and by 19 \% in both Tersoff-contact runs, while the band narrows to 4 to 14 \% of the median at the last iteration: the particles then no longer sample the search space independently, but move together around a common, and possibly local, minimum. The swarm also learns quickly where a structure can be built: the inadmissible candidates, whose construction failed or whose properties fell too far from a target to be scored, a quarter of the initial swarm in run Q-H and half in run Q-T, disappear from the third and the fourth iteration onward, respectively, and the synthetic-target runs never produced one. The candidates proposed by the quadratic model never improved on the best of the swarm, so that the convergence reported here is that of the swarm alone.

The lowest best cost is reached in run Q-H, 0.470, and the highest in run S-T, 0.738. In the three runs other than S-T most of this residual comes from the local RDF, 73 to 82 \% of the best cost, while the density, the surface area and the PSD together account for the rest (Supplementary Note 2). This is intended: the construction places the global geometry where it is asked to, whereas the local order, set during growth by a placement potential which the anneal later replaces, is approached but not reached. As expected, the synthetic target is harder to reach on its global terms: its two PSD terms make up 25 \% of the best cost of run S-H and 50 \% of that of run S-T, against 9 \% in run Q-H, although the synthetic-target runs are scored on one property fewer, consistent with the difficulty of its small mode.

The noise of the construction sets how much of this convergence can be resolved. Two structures built from the same parameters with different random seeds differ, and so do their costs: the standard error of a mean cost over seven replicates is 0.018 to 0.024, i.e., 3 to 5 \% of the best costs (Supplementary Note 2), and two candidates closer than about two standard errors cannot be ranked. Against this scale the search is real: the best candidate improves on the best of the random initial swarm by 2.5 to 6.3 standard errors. No single step of the best cost exceeds about two standard errors, so that this gain is a sum of steps each within the noise, which the runs accumulate differently. Run S-H gathers all of it within three iterations and run Q-H more than half, its later changes adding 1.5 standard errors in total. Run S-T finds its best candidate at the ninth iteration and never improves on it. Run Q-T, instead, keeps gaining until about its fifteenth iteration, as does its median (Figure \ref{fig:results_pso}A), its later changes adding 1.5 standard errors, within the noise. Within two standard errors of the best lie 5 \% of the candidates of run Q-H, 7 \% of Q-T, 19 \% of S-H and 16 \% of S-T, the large share of run S-H marking a flat basin, in which the swarm keeps finding solutions equivalent to its best and none better. The recovered parameter vector is therefore one representative of a set of statistically equivalent solutions rather than a unique optimum, and which of its components the cost actually responds to is what the history of the search can tell.

\subsubsection{The search history shows which parameters the cost responds to}

Convergence shows that the swarm found a region of low cost, but not which of the searched parameters mattered. The history of each search can be read as a sample of the cost surface at no additional construction cost, first by inspection and then through a model of the cost. A similar reading was recently applied to swarm calibrations of battery manufacturing models \cite{vijayaraghavan2026accelerated}. A swarm, however, visits a region both because its cost is low and because the particles are drawn toward their best positions, so that the positions it sampled do not by themselves show what the cost responds to.

Figures \ref{fig:results_pso}B and \ref{fig:results_pso}C project the sampled cost onto one parameter at a time, each candidate colored by its iteration and the recovered values marked by dashed lines. Along $\phi$ in run Q-H the candidates draw a clear parabola, the late ones gathering at its vertex: the cost responds strongly to how much pore volume is placed. The range $B$ of the same run converges as clearly, but along a shallower and blurred parabola, since its best value depends on $\phi$. Figure \ref{fig:results_pso}C shows instead two ways in which a parameter escapes determination: the depth $A$ of run Q-T drifts toward its lower bound, although its best candidate sits a quarter into the range, and the range $B$ of run S-T piles up against its upper bound, a third of its candidates sitting at 2.5 \AA{}, so that there the search range, and not the data, sets the recovered value.

Such projections suggest which parameters matter but do not prove it: a swarm contracts along every parameter, whether or not the cost depends on it, so that the lowest costs appear next to the recovered value along every axis (Supplementary Note 3). Separating what the cost responds to from what the swarm merely carried along requires a model of the cost. The model used here is a second-order polynomial of the searched parameters, the lowest order able to represent an interior optimum, fitted by least squares to every candidate after the random initial swarm (see Section \ref{sec:pso} and Supplementary Note 3). It captures most of the cost variance in the harmonic runs ($R^2$ = 0.89 in run Q-H and 0.85 in S-H) and less in the Tersoff-contact runs (0.70 and 0.64). The replicate noise, which no fit can capture, accounts for at most a tenth of the cost variance (see Section \ref{sec:pso}), so that the 30 to 36 \% of the variance which the fit misses in the Tersoff-contact runs is not noise, but a shape of the cost surface which a quadratic cannot follow.

The share of the cost variance which each parameter explains alone, its drop-one partial $R^2$, is the loss of $R^2$ when the model is refitted without that parameter (Figure \ref{fig:results_pso}D). It is consistent across the four runs: $\phi$ is the most informative parameter, explaining a quarter to three quarters of the variance, $B$ is second, and $k_\varphi$, $A$ and $m$ explain 5 \% or less each (Supplementary Note 3). This share is conservative, since variance which two parameters explain together, because they moved together along the trajectory, is credited to neither, the remaining parameter taking over the role of the one removed: it is robust to a parameter which merely drifts while the cost decreases, but it under-credits one which moves in step with another, as $B$ in run S-H.

\subsubsection{The cost is stiff along the pore parameters and sloppy along the others}\label{sec:curvature}

The curvature of the fitted surface describes how the cost responds to combinations of parameters. Being quadratic, the fit has a single Hessian matrix for the whole sampled region, whose eigenvalues, the principal curvatures, measure how fast the cost rises when the parameters move away from the optimum along each eigenvector, in either sense. Along a stiff direction a small move raises the cost markedly, and the data fix that combination, whereas along a sloppy direction the cost barely changes and the data leave it free \cite{transtrum2015perspective}. A negative curvature marks a direction with no interior minimum, along which the search is pushed toward a bound. A parameter is classified as stiff when it explains at least 5 \% of the cost variance alone and dominates, i.e., carries the largest component of, a direction of positive curvature, the attribution being robust when the same parameter dominates that direction in at least 85 \% of the resamples of the particle trajectories (see Section \ref{sec:pso}). A run is taken as identified when at least one parameter is stiff with a robust attribution. Even with only sixteen particles per run, the resampling tells a stiff direction from a sloppy one, although its intervals are too wide to separate two directions of similar stiffness.

Figure \ref{fig:results_pso}E plots the eigenvalues of each run in decreasing order, so that the direction along which the cost responds most comes first, with their 95 \% intervals and a filled marker for a robust attribution, and gives a clear picture for three of the four runs (Supplementary Note 3). In runs Q-H, S-H and Q-T two directions stand above the rest, the stiffest dominated by $\phi$ in at least 85 \% of the resamples and the second by $B$, but only $\phi$ is stiff under both targets and both placement potentials. The attribution of $B$ is robust only in run Q-H, and in runs S-H and Q-T the second curvature lies closer to its neighbors than in run Q-H. Two close curvatures leave their directions poorly defined: the cost rises almost equally fast along either of them and along any combination of the two, as in a bowl of nearly circular section, so that a small change of the sampled history is enough to rotate one direction into the other. In both runs $B$ therefore dominates its direction in only 60 \% of the resamples, a real but weaker second parameter. In run S-H this is the flat basin met in its convergence: the swarm stopped improving early because, beyond $\phi$, the cost hardly distinguishes between the candidates it kept finding.
Figure \ref{fig:results_pso}F gives the composition of the two stiffest directions of each run, i.e., of the eigenvectors of the two largest eigenvalues of Figure \ref{fig:results_pso}E, as signed components with the sign chosen so that $\phi$ is positive, and shows why $\phi$ and $B$ are hard to separate. The stiffest direction is made almost entirely of $\phi$ and $B$ with the same sign in every identified run, and the second of the two with opposite signs. In the plane of these two parameters the cost forms a long, narrow valley, whose floor runs along the diagonal on which one parameter grows as the other shrinks: a step across the valley, raising or lowering $\phi$ and $B$ together, raises the cost three to six times more than a step of the same length along its floor, the ratio of the first two eigenvalues in Figure \ref{fig:results_pso}E. Both parameters define the void the carbon leaves, $\phi$ through the pore volume placed and $B$ through the standoff, so that walking along the floor trades pore volume for standoff, which also blurs the parabola of $B$ in Figure \ref{fig:results_pso}B. The remaining parameters, $k_\varphi$, $A$ and $m$, share the nearly flat directions while being recovered inside their ranges: the cost is indifferent to them, which marks them as sloppy rather than well chosen. For the pore wall this split is expected. Since the repulsion of the wall scales as $d^{-12}$, the standoff grows only as $A^{1/12}$, so that doubling $A$ moves it by about 6 \%, while $B$ sets it directly. More revealing is the sloppy dihedral stiffness: once the pores are in place, the stiffness of the sheets barely affects the global geometry of the targets.

In run S-T the early plateau is unidentifiability rather than convergence. Its fitted cost is the poorest (lowest $R^2$), no curvature dominates, neither $\phi$ nor $B$ dominates a direction in more than two thirds of the resamples, and both settle at bounds, $B$ at 2.5 \AA{} and $\phi$ at 0.04. The cause is a conflict between two targets under the Tersoff-contact construction: placing the bimodal pore volume raises the density error faster than it lowers the PSD error (Supplementary Note 3), so that the search lowers its cost by placing almost no pores. Removing the conflict confirms it: when the same candidates are ranked with the density term removed from their cost, the best of them moves off both bounds, to $\phi$ = 0.24, while the harmonic construction, which reaches the target density at $\phi$ = 0.26 in run S-H, shows no such conflict. The failure therefore belongs to the Tersoff-contact construction for this target, not to the search, and it can be determined from the search history alone, without an additional construction.

For these two targets, $\phi$ is constrained in the three identified runs and $B$ only in run Q-H, whereas $k_\varphi$, $A$ and $m$ are not, so that the search could be reduced to the two pore parameters. The same reading also checks the result of a search, since the identification criterion flags run S-T, the one whose target could not be reached, from its history alone. How closely the recovered structures reproduce each target is examined next.

\subsection{The global geometry is recovered, local order requires annealing}\label{sec:recovered}
\begin{figure}
    \centering
    \includegraphics[width=1\linewidth]{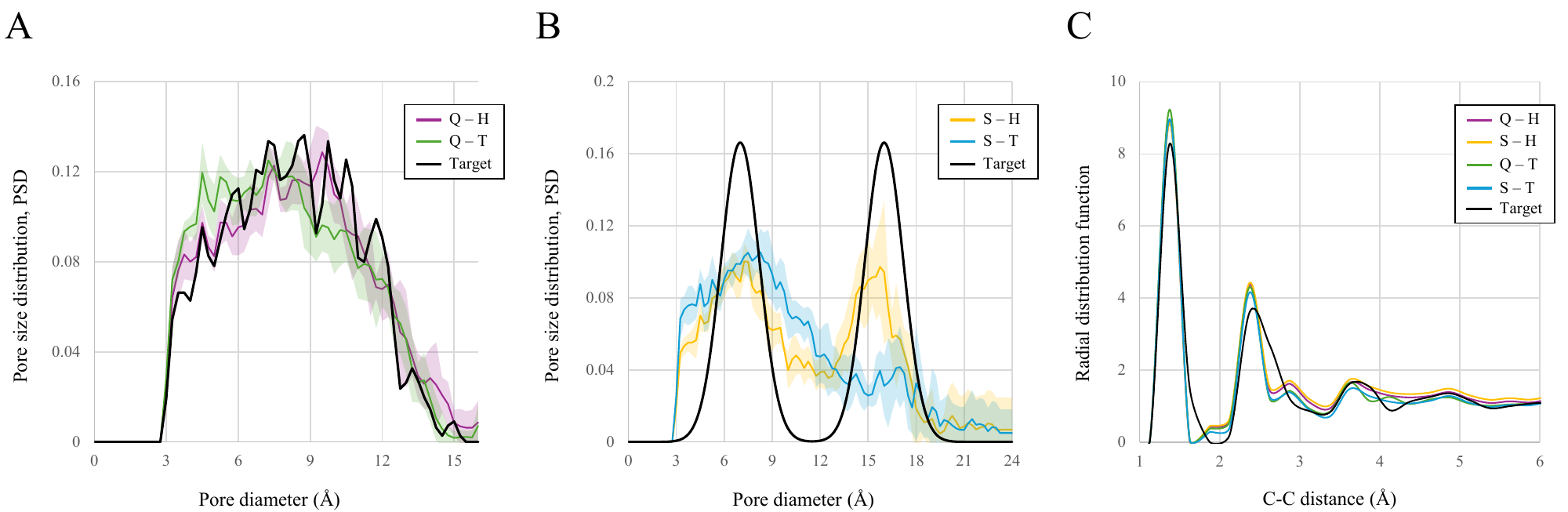}
    \caption{\textbf{Properties of the recovered structures against their targets.} (A) Pore size distribution of the quench-MD target (black) and of the structures recovered under the two placement potentials, runs Q-H and Q-T. (B) Same PSD comparison for the synthetic bimodal target, whose two modes of equal volume lie at 7 \AA{} and 16 \AA{}, and runs S-H and S-T. (C) Radial distribution function of the four proxy structures and of the quench-MD reference (black), computed over the whole box in 0.25 \AA{} bins and plotted at the bin centers. In every panel the solid line is the mean over the seven replicates of each candidate and the band spans one standard deviation on either side of it.}
    \label{fig:results_anchors}
\end{figure}
The total cost alone does not show how closely each property matches its target. The seven replicates of each optimal parameter set are therefore compared with the targets property by property (Table \ref{tab:results_scalar}), first on the scored properties and then on three properties that were not scored, namely the probe-occupiable and the geometric pore volumes and the skeletal density, measured on the QMD reference with the same routines. The synthetic target prescribes only a density and a PSD, so that runs S-H and S-T are judged against it on these two properties alone, and on the unscored ones can only be compared with the Q runs. Except for the density of run Q-T, the Q runs reproduce the scored global properties of their target, and the probe-occupiable pore volume, which was not scored, while run S-H recovers the bimodal shape of the synthetic target and run S-T does not. In all four runs, however, the local order, measured by the RDF and the coordination, departs from that of the reference.

\begin{table}[h]
\centering
\caption{Properties of the optimal proxy structures against their targets (with the QMD labeled as 'Q', and the synthetic as 'S') as mean over seven replicates of each candidate. The synthetic target prescribes only the density and the PSD. The reference values of the unscored properties are those of the 100 \AA{} QMD structure, measured with the same routines and the same probe parameters. Unlabeled targets refer to the QMD reference.} \label{tab:results_scalar}
\begin{tabular}{@{}lllllll@{}}
\toprule
Property & Scored & Q-H & Q-T & S-H & S-T & Target\\
\midrule
Density (g cm\textsuperscript{-3}) & \checkmark & 0.942 & 0.904 & 0.946 & 0.934 & 0.950 (Q \& S) \\
Accessible Surface Area (m\textsuperscript{2} g\textsuperscript{-1}) & \checkmark & 1528 & 1591 &  1394 & 1186 & 1580\\
Probe-occupiable pore volume (cm\textsuperscript{3} g\textsuperscript{-1}) & $\times$ & 0.493 & 0.499 & 0.504 & 0.412 & 0.518\\
Geometric pore volume (cm\textsuperscript{3} g\textsuperscript{-1}) & $\times$ & 0.609 & 0.642 & 0.612 & 0.607 & 0.620\\
Skeletal density (g cm\textsuperscript{-3}) & $\times$ & 2.21 & 2.15 & 2.25 & 2.15 & 2.31\\
2-coordinated atoms (\%) & $\times$ & 49.4 & 48.9 & 49.4 & 49.5 & 12.2\\
3-coordinated atoms (\%) & $\times$ & 50.6 & 50.2 & 50.6 & 50.3 & 80.9\\
\midrule
\end{tabular}
\end{table}

On the scored properties, the harmonic runs reach the target density to within 1 \%, and run Q-H the surface area to within 3.3 \%. The largest deviation is that of run Q-T, which matches the surface area but stays 5 \% below the target density, its search trading one channel for the other (Supplementary Note 4). The PSD of the QMD target is reproduced closely under both placement potentials (Figure \ref{fig:results_anchors}A), with an earth-mover distance (EMD) of 0.12 \AA{} in run Q-H and 0.18 \AA{} in run Q-T. The EMD is the mean distance, in pore radius, over which the pore volume of one distribution has to be moved, toward smaller or larger pores, to obtain the other (see Section \ref{sec:costing}), so that these values are 3 and 4 \% of the mean pore radius of the target, the scale on which the cost scores them. The bimodal target instead separates the two placement potentials (Figure \ref{fig:results_anchors}B). Run S-H recovers both modes at the intended diameters, with 56 \% of its pore volume in the smaller one against 50 \% in the target and an EMD of 0.57 \AA{}, 10 \% of the mean pore radius of the target; its peaks are lower than those of the target, part of the volume spreading between and beyond the modes, but the bimodal shape is recovered. Run S-T holds 69 \% of its pore volume below the 11 \AA{} valley between the modes, but with seven times fewer placed pores than run S-H, so that its small mode is made of the interstitial voids of the carbon network (Supplementary Note 5). How much the placement potential matters therefore depends on the target. For the QMD target the two constructions recover the PSD comparably, and the Tersoff-contact one misses the target density by 5 \%, whereas for the bimodal target only the harmonic construction places the pore volume at the target density, the Tersoff-contact construction approaching that density by placing almost no pores (Section \ref{sec:curvature}).

The same interstitial voids set a lower limit to the pore sizes the method controls. Every proxy structure holds 5 to 8 \% of its pore volume below 4 \AA{}, the two bimodal runs included, whose target places only 0.3 \% there. Since the diameter is that of the largest empty sphere measured from the atom surfaces, the atoms bounding a 4 \AA{} pore sit about 7.4 \AA{} apart, slightly more than twice the 3.35 \AA{} interlayer spacing of graphite: below this size the PSD is set by the carbon network rather than by the placed pores, and no searched parameter changes it. For the QMD target, which itself holds 5.9 \% of its pore volume in that range, this limit costs little, whereas for a target without micropores it is a resolution limit of the construction.

Of the unscored properties, the probe-occupiable pore volume is the one which depends on the shape and the connectivity of the void, and not only on its amount: it is the volume a nitrogen probe can occupy while reaching it from the box faces (see Section \ref{sec:post-process}). Runs Q-H and Q-T reproduce it to within 5 \% of the reference. The skeletal density, i.e., the density of the carbon network without its pores, is instead lower than that of the reference in every run, and is set by the placement potential rather than by the target: 2.21 and 2.25 g cm\textsuperscript{-3} under the harmonic potential, and 2.15 g cm\textsuperscript{-3} in both Tersoff-contact runs. The Tersoff-contact construction, which has neither the dihedral term that holds a new ring in the plane of its parent nor the attraction that draws open edges together (see Section \ref{sec:method_potential}), thus builds the sparsest network. The density missed by run Q-T goes into this sparser network rather than into pore volume, since its geometric pore volume, measured without a probe and without requiring the void to be connected, stays close to that of the reference.

The local order shows where the proxy structures depart from the reference (Figure \ref{fig:results_anchors}C and Table \ref{tab:results_scalar}). The first two peaks of every RDF fall at the bond length and at the second-neighbor distance of the reference, but are higher. The construction places each ring as a rigid template at the equilibrium bond length and angle, so that first and second neighbors sit at single distances, whereas the thermalized reference has broader shells. The topology is incomplete: almost half of the atoms are two-coordinated, against 12 \% in the reference, marking where a growth front, an open edge to which new rings attach, stopped, and four-coordinated atoms are essentially absent. Since the coordination is nearly the same in the four structures, it is set by the construction rather than by the target or the placement potential. This is the local term which dominates the remaining cost of the searches (Section \ref{sec:convergence}), and it is left to the anneal rather than to a refinement of the search.

The construction therefore yields a network that is consistent in geometric properties, yet incomplete in topology. The next section investigates whether annealing recovers the local properties without compromising the global structure.

\subsection{Annealing improves local coordination while retaining the main features of the pore structure}\label{sec:anneal}

Section \ref{sec:recovered} left the proxy structures with the global geometry of their targets but with an incomplete local topology: almost half of their atoms are two-coordinated, and four-coordinated atoms are essentially absent. The anneal of Step 5 is meant to complete the network while leaving the pores in place. Three replicates of the optimal proxy structure of each run were annealed with EDIP in four stages, a 1 ps burn-in ramping from 300 K to 2500 K, a 20 ps hold, a quench to 300 K and a short cold hold, 39.1 ps in total (see Section \ref{sec:method_anneal}), the pores being kept as repulsive walls along the whole trajectory. To separate the effect of the protocol from that of the construction, the QMD reference was annealed with the same protocol, without walls. The reference barely changes: its share of three-coordinated atoms moves from 80.9 to 81.7 \%, its pore volumes and surface area by less than 1 \% and its mean pore diameter by 0.14 \AA{}, while its ring ratio, the number of five- and seven-membered rings per six-membered ring, relaxes from 1.10 to 1.05 (Supplementary Note 6). The changes of the proxy structures discussed below are therefore due to the defects left by the construction, namely the open growth fronts of Section \ref{sec:recovered}, and not to the protocol.

Handing a constructed structure to the dynamics requires two precautions, one removing the heat of the cure where it is released, the other keeping the walls from binding carbon to the pore surfaces. First, closing the growth fronts is exothermic: during the 1 ps burn-in the proxy structures release 0.10 to 0.15 eV per atom more than the reference heated along the same ramp. The burn-in therefore uses a Langevin thermostat, which damps each atom independently with a damping time of 0.01 ps and so removes the heat where it is released, keeping the temperature within 35 K of the ramp after the first 0.1 ps. Second, the pore walls are cut at their own minimum. Carried into the anneal in full Lennard-Jones form, a wall would keep an attractive well which, depending on the depth and the range recovered by the search, can be comparable to or deeper than the thermal energy at 2500 K (Supplementary Note 7), so that it would bind carbon to the pore surfaces rather than only keep the pores empty. Cut at its minimum, it retains only its repulsive branch. A third effect is specific to the software used. The strain of the construction, including the atoms pressed against the walls, is not released by the initial minimization (see Section \ref{sec:method_anneal}) and comes out within the first 0.1 ps as a brief temperature excursion, to 820 to 1060 K in runs Q-H, S-H and S-T and to 2680 to 2960 K in run Q-T, close to the glass transition reported for EDIP \cite{best2020evidence}. Relaxing the proxy structures of runs Q-H and Q-T before the dynamics removes this excursion and leaves the final structures unchanged within the spread between replicates, except for a slightly lower ring ratio (Supplementary Note 7), so that the results reported below do not depend on this excursion.

\begin{figure}
    \centering
    \includegraphics[width=1\linewidth]{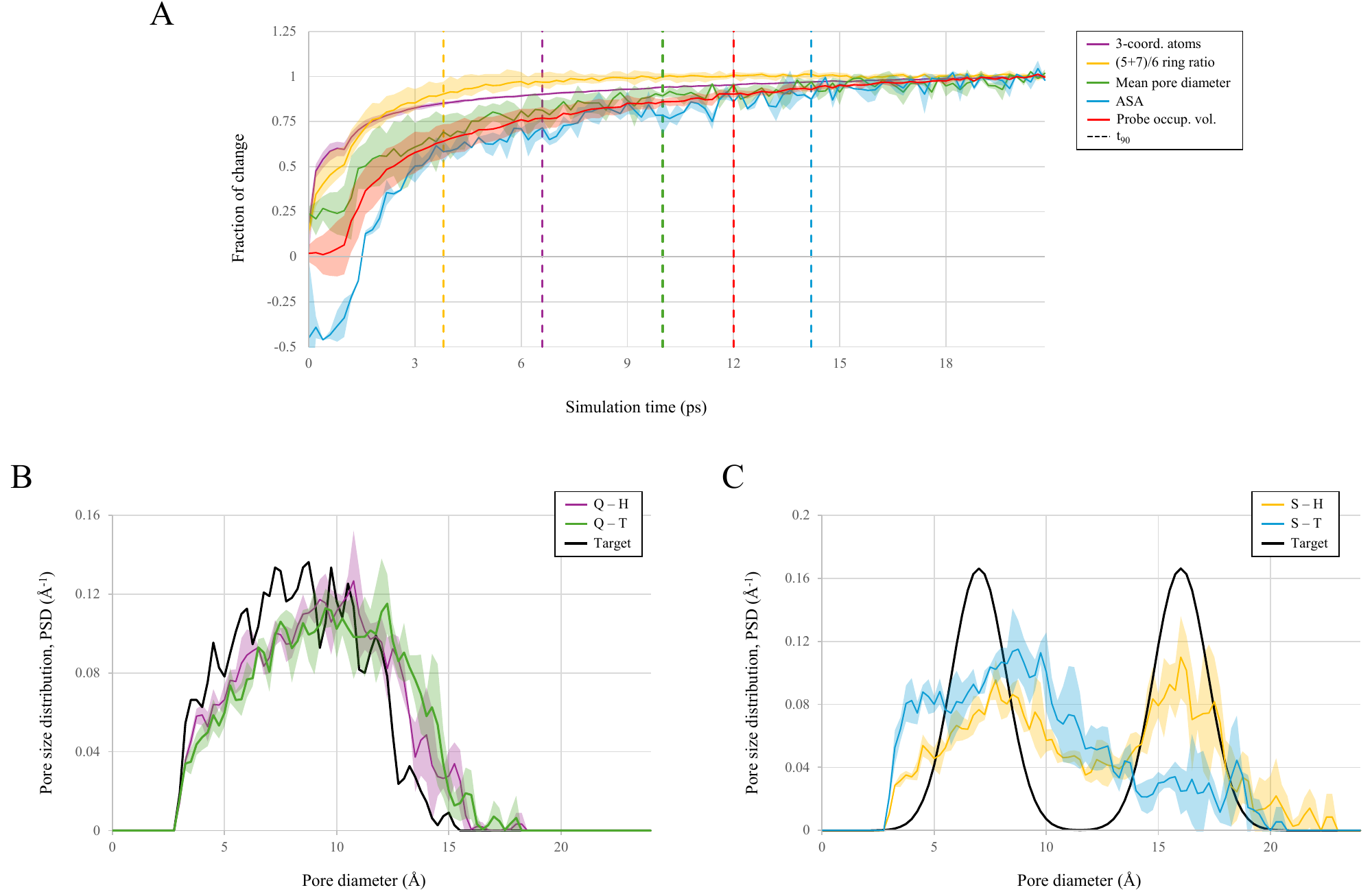}
    \caption{\textbf{Convergence of the anneal and pore size distributions after it.} (A) Fraction of the change completed against simulated time during the burn-in and the hold, from the proxy structure (0) to the end of the hold (1), for the three-coordinated share and the ring ratio of the four runs and for the mean pore diameter, the accessible surface area and the probe-occupiable volume of the Q runs; mean over the runs, the band spanning their range. Negative values mark an initial change opposite to the overall one. Vertical dashed lines: convergence time $t_{90}$ of each mean curve, i.e., the time at which it completes 90 \% of its change. (B, C) Pore size distribution after the anneal of the runs targeting the QMD structure (B) and the synthetic bimodal PSD (C), against their targets; the distributions before the anneal are those of Figure \ref{fig:results_anchors}A and B. Solid lines: mean over three replicates; bands: one standard deviation.}
    \label{fig:anneal_dynamics}
\end{figure}

Figure \ref{fig:anneal_dynamics}A follows the anneal through the fraction of the change that five properties have completed, from the proxy structure to the end of the hold. Convergence is judged on these fractions rather than on the energy, which below the glass transition keeps decreasing slowly, and also falls for the reference annealed with the same protocol (Supplementary Note 6). The local topology converges first: about half of the change in coordination occurs during the 1 ps burn-in, the time to complete 90 \% of it, $t_{90}$, being 6 to 7 ps, and the ring ratio reaches 90 \% of its change within 2 to 4 ps. The pore network of the Q runs is slower, its mean pore diameter, accessible surface area and probe-occupiable volume reaching 90 \% of their change within 8 to 14 ps, so that the length of the hold trades the cure of the topology against the shift of the pores. Every $t_{90}$ thus falls within the hold, after the burn-in ramp, at a constant 2500 K, where the share of the bonds that thermal motion stretches beyond the fixed bond cutoff does not change from frame to frame. Over the last 5 ps of the hold every quantity changes by less than 10 \% of its whole change, and the annealed structures are compared below on their final frame at 300 K.

\begin{figure}
    \centering
    \includegraphics[width=1\linewidth]{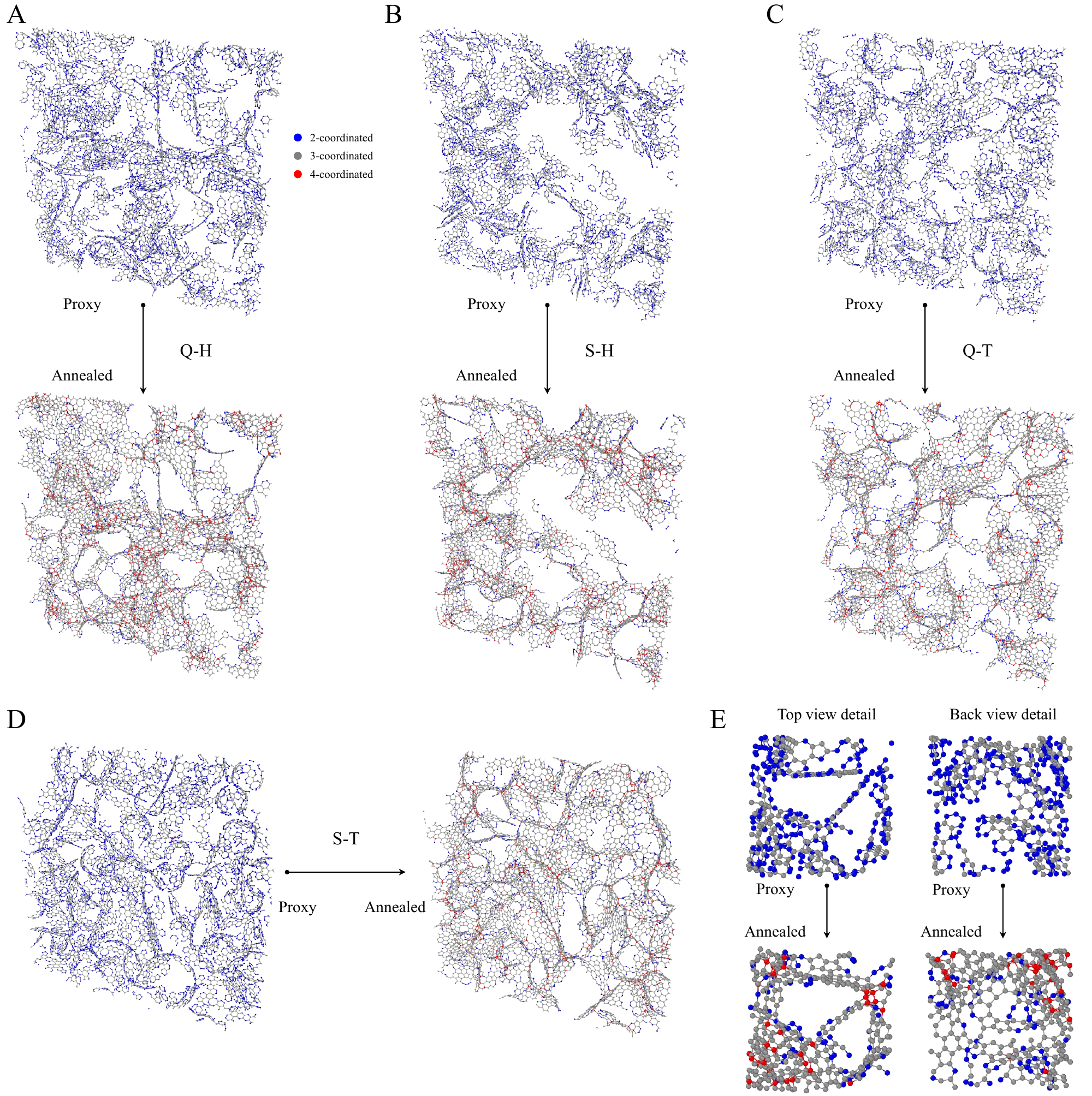}
    \caption{\textbf{Proxy structures before and after the anneal.} (A-D) Thin slabs (10 \AA{}) comparing the optimal proxy structures with the same slabs after the anneal of runs Q-H (A), S-H (B), Q-T (C), and S-T (D), with the atoms colored by their coordination (two-coordinated in blue, three-coordinated in grey, four-coordinated in red). (E) Close-ups of a single pore of a structure resulting from the Q-H PSO run, (20$\times$20$\times$20 \AA{}\textsuperscript{3} cubic slice) comparing the atoms organization and coordination before (top row) and after (bottom row) annealing.}
    \label{fig:anneal_ovito}
\end{figure}

Figure \ref{fig:anneal_ovito} compares a thin slab of each proxy structure with the same slab after the anneal. Before the anneal, the four proxy structures share the same features: the growth fronts, highlighted by the two-coordinated atoms (blue), are spread across the whole slab, and the carbon network is often fragmented, a direct consequence of placing each ring as a rigid template. Run S-H (Figure \ref{fig:anneal_ovito}B) shows large voids next to regions of smaller pores, as its bimodal target requires, while the Tersoff-contact construction produces a sparser network with sporadic chains of atoms, consistent with its lower skeletal density (Table \ref{tab:results_scalar}). In run S-T, which places only a few pores, the construction wall seats the carbon rings on the pore surfaces as nearly spherical shells (Figure \ref{fig:anneal_ovito}D), most of the void being left as interstitial space in the carbon network. After the anneal, the pores are still in place, whereas the carbon fragments are smoother and bent, far fewer two-coordinated atoms remain, and the four-coordinated atoms (red) appear mainly where fragments join (Figure \ref{fig:anneal_ovito}E). A similar morphology, i.e., curved graphene-like fragments rather than flat sheets, is reported for QMD and EDIP models of carbide-derived carbons at comparable densities \cite{palmer2010modeling,de2017structural}, where negative curvature is found to dominate \cite{martin2019topology}.

\begin{table}[h]
\centering
\caption{Proxy structures before and after the anneal, against the 100 \AA{} QMD reference. Ranges over the four runs, or over runs Q-H and Q-T for the pore network, whose reference applies to the Q runs only, each run being the mean of three replicates. Coordination and rings are measured on the final frame at 300 K. With the settings used for the annealed structures, the reference has a surface area of 1598 m\textsuperscript{2} g\textsuperscript{-1}, 1.1 \% above the target of 1580 m\textsuperscript{2} g\textsuperscript{-1} scored by the search (Table \ref{tab:results_scalar}). The values of each run, with the spread between replicates, are reported in Supplementary Note 8.} \label{tab:anneal}
\begin{tabular}{@{}llll@{}}
\toprule
Property & Before & After & Reference\\
\midrule
2-coordinated atoms (\%) & 49.0--49.5 & 11.4--13.7 & 12.2\\
3-coordinated atoms (\%) & 50.2--50.6 & 77.8--81.7 & 80.9\\
4-coordinated atoms (\%) & 0.01--0.05 & 5.44--10.12 & 6.92\\
(5+7)/6 ring ratio & 0.50--0.52 & 0.87--0.91 & 1.10\\
RDF distance to the reference & 0.212--0.299 & 0.037--0.201 & --\\
Skeletal density (g cm\textsuperscript{-3}) & 2.15--2.25 & 2.30--2.35 & 2.31\\
Pore connectivity $k_\mathrm{avg}$ & 1.745--1.787 & 1.627--1.636 & 1.629\\
Mean pore diameter, Q runs (\AA{}) & 7.98--8.34 & 9.12--9.47 & 8.30\\
Geometric pore volume, Q runs (cm\textsuperscript{3} g\textsuperscript{-1}) & 0.612--0.639 & 0.634--0.672 & 0.620\\
Accessible Surface Area, Q runs (m\textsuperscript{2} g\textsuperscript{-1}) & 1522--1594 & 1577--1650 & 1598\\
\midrule
\end{tabular}
\end{table}

The local topology is cured in all four runs (Table \ref{tab:anneal}). The share of two-coordinated atoms, i.e., of the open growth fronts, falls from almost half to 11 to 14 \%, around the 12 \% of the reference, and the share of three-coordinated atoms approaches it, while the RDF distance from the reference falls in every run, as the thermal motion broadens the neighbor shells that the rigid templates had set at single distances. The only clear difference between the runs is the share of four-coordinated atoms, 9 to 10 \% with the harmonic placement potential against 5 to 8 \% with the Tersoff-contact one: since the protocol is the same, this difference is inherited from the construction.

The anneal also compacts the sparse network left by the construction, bringing the skeletal density, which was not scored, to that of the reference in every run. Since the volume of the box is fixed, the volume freed by the compaction goes to the pores, and the pore size distribution shifts toward larger pores. In the Q runs, whose target is the QMD structure itself, the mean pore diameter grows by 0.8 \AA{} in run Q-H and by 1.5 \AA{} in run Q-T (Figure \ref{fig:anneal_dynamics}B, to be compared with Figure \ref{fig:results_anchors}A), while its width and the share of the pore volume in the smallest pores stay close to those of the target. The distance from the target therefore grows through a shift of the distribution rather than a change of its shape, and the pores widen at a nearly constant pore volume: the geometric pore volume and the surface area grow by a few percent only, and stay within 3 \% of the reference in run Q-H and within 9 \% in run Q-T. Part of the shift is set by the walls. The construction accepts a ring with a positive wall energy below the threshold, so that the proxy structures enter the anneal partly within the range of the walls, which push these atoms out at the beginning of the burn-in. This accounts for about a tenth of the shift of the mean pore diameter in run Q-H and almost half in run Q-T, whose walls carried the largest energy, within the first picosecond. At the end of the anneal the walls hold less than 0.002 eV per atom, so that the rest of the shift, which develops along the hold, is due to the compaction of the carbon network, and the annealed structures do not rely on the walls to keep their pores open. The pore connectivity, i.e., the probe-accessible links per pore of the void network (see Section \ref{sec:post-process}), converges to that of the reference as well, in all four runs.

The runs targeting the synthetic PSD have no reference structure for their global properties, which can only be judged against their own targets (Supplementary Note 8). Run S-H keeps its properties, its PSD moving toward the target, as the comparison of Figure \ref{fig:anneal_dynamics}C with Figure \ref{fig:results_anchors}B shows, its distance from it falling by a fifth. Run S-T, which the search left unidentified (Section \ref{sec:curvature}), changes the most: its accessible surface area increases by 28 \% and its probe-occupiable volume by 25 \%, while its PSD and its geometric pore volume barely move. The anneal thus appears to open voids which were already there rather than to widen them: the shells which the construction seated on the pore surfaces (Figure \ref{fig:anneal_ovito}D) enclose part of the void, which the PSD counts but the accessible measures, reached from the box faces, do not, until the anneal breaks the shells.

Among the untargeted properties, the ring statistics change the most, the ring ratio increasing from about 0.5 to about 0.9, against 1.10 in the reference. Only a quarter of the rings formed during the anneal are hexagonal, and about half are five- or seven-membered: closing the growth fronts of fragments joined at sharp angles produces mostly odd-membered rings, which accommodate the curvature visible in Figure \ref{fig:anneal_ovito}. The origin of the remaining gap is not resolved here: it may reflect the ring probabilities of the construction (see Section \ref{sec:structure_generation}), or the limited mobility of the network below the glass transition reported for EDIP \cite{best2020evidence}.

In terms of simulated time, the anneal of a proxy structure converges within about 15 ps at 2500 K, and the whole protocol lasts 39.1 ps. For comparison, the QMD protocol used to produce the reference takes 370 ps (Section \ref{sec:qmd_protocol}), and the quench and anneal protocols reported for carbide-derived carbons last up to 500 ps with ReaxFF \cite{thompson2017atomistic} and 2 to 10 ns with EDIP \cite{de2017structural}. These are time scales rather than a comparison of performance, since the construction is preceded by the search of Step 4. The search, however, is paid once per target, and every further structure for that target costs one construction and one anneal.

The construction therefore sets the global geometry of the structure, namely the position, the shape and the width of its pore size distribution, while the anneal completes its connectivity, recovering the coordination of the reference. The anneal also compacts the carbon skeleton to the skeletal density of the reference, which, at fixed density, shifts the pore size distribution toward larger pores while keeping its shape.

\section{Conclusions}

This work presents a generator of nanoporous carbon structures which takes as inputs the pore size distribution and the volume fraction of the pores to be placed, together with an inverse-design procedure that recovers the generation parameters from a target structural fingerprint. Pores are placed first, sampled from a tabulated distribution of arbitrary shape, and a carbon network is grown ring by ring around them; a particle swarm then searches the generation parameters against the target, and a short anneal by molecular dynamics completes the network.

First, the search reproduces the scored global properties of a reference structure produced by quenched molecular dynamics, its density and surface area within about 5 \% and the shape of its pore size distribution, together with its probe-occupiable pore volume, which was not scored. As a stress test, the harmonic construction also recovers both modes of a synthetic bimodal distribution, which no other route has been shown to reach, whereas the Tersoff-contact construction, in which the Tersoff energy acts only on contacts with distant parts of the network and so carries no local bonding term, places almost no pores for the same target, which shows that the role of the placement potential depends on the target. Then, a response surface fitted to the history of the search separates the parameters that the targets actually constrain, the pore volume fraction in every identified run and the range of the pore wall in one, from the sloppy ones, to which the cost is indifferent, and distinguishes a converged search from a stalled one without any further construction. Finally, a short anneal, not optimized for this work, is enough to cure the coordination of the proxy structures and to bring their skeletal density and pore connectivity to those of the reference without targeting them, while the pores widen at a nearly constant pore volume, the pore size distribution keeping its shape and its mean diameter growing by 0.8 and 1.5 \AA{} in the two runs targeting the quenched structure. The walls hold almost no energy at the end of the anneal, so that the annealed structures keep their pores without them. Every structural property converges within about 15 ps of simulated time, in a protocol about ten times shorter than the quench which produced the reference. Molecular dynamics still supplies the local chemistry, and the ring statistics remain below those of the reference, a gap whose origin is not resolved here.

The method thus turns the pore geometry from an outcome of the simulation protocol into a specification. The search, affordable because the cost of measuring the pore size distribution of each construction grows only linearly with the size of the structure, is paid once per target, and every further structure for that target requires one construction and a short anneal in place of a full quench trajectory, for elemental carbon in the present form of the construction.

Future improvements include targets taken directly from adsorption measurements, in place of the simulated and synthetic targets tested here, multimodal and mesoporous distributions in larger boxes, and searches designed for the identifiability analysis as well as for the optimum, with more particles and fewer iterations at the same number of constructions. In the longer term, building a carbon around a prescribed void geometry turns the design of electrodes and sorbents into a question that can be investigated in silico, for a given electrolyte or adsorbate rather than for a given synthesis route.

\section{Methods}

%The last part of this section describes the reference data (Sections 6 and 7) and the validation setup of the post-processing routines (Section 8). 
%The algorithmic details of every step are reported in the Supporting Information.

\subsection{Structure generation}\label{sec:structure_generation}

Structures are built by explicit pore pre-conditioning: the void is placed first, and carbon is grown around it. 

\subsubsection{Pore placement}

Pore radii are drawn from the target pore size distribution (PSD), supplied either as a parametric Gaussian distribution or as a tabulated curve of arbitrary shape, and spheres are inserted sequentially into the cell, periodic along every axis in this work, until a pre-determined pore volume fraction $\phi$ is reached. Spheres are allowed to overlap, since two intersecting spheres simply share the void they define.
A tabulated PSD is, by the convention of adsorption and pore-network analysis and by that of the measurement described in Section \ref{sec:post-process}, a differential volume density $dV/dD$ \cite{thommes2015physisorption}. Since pores are placed one at a time, namely by count, and a pore of diameter $D$ contributes a volume proportional to $D^3$, the count density $n(D)$ reproducing a target volume distribution is:
\begin{equation}
    n(D) = \frac{1}{D^{3}C}\frac{\mathrm{d}V}{\mathrm{d}D} \,, \qquad C = \int \frac{1}{D^{3}}\frac{\mathrm{d}V}{\mathrm{d}D}\mathrm{d}D \, , 
\end{equation}
where $C$ normalizes $n(D)$ to unit integral. The sampler therefore divides the tabulated curve by the pore volume before inverting the cumulative distribution; without this step, a mode carrying a large fraction of the volume in comparatively few pores is under-placed by volume. The complementary convention, in which the tabulated curve is used as $n(D)$ directly, is also implemented, since pore counting from imaging techniques returns a PSD by count. % Which of the two a given input follows is a declaration about the data and not a tunable parameter, so it is stated together with each target.

The placed spheres are exclusion zones, which no carbon atom may enter. Since the carbon growth changes the void (see Section \ref{sec:method_overview}), all pore quantities reported in this work are measured on the final configuration (Step 2, see Section \ref{sec:post-process}), and $\phi$ is a parameter of the generation, not the porosity of the final structure.

\subsubsection{Ring-by-ring carbon growth} 

Carbon is grown as a sequence of rings, so that every atom carries two or three bonds by construction. The construction aims at maximizing the three-coordinated, sp\textsuperscript{2} sites, the two-coordinated atoms marking where a growth front has not closed, a defect of the proxy structure rather than a feature of the final material. An initial ring is seeded and its atoms become anchors, namely the unsaturated sites at which a further ring may be attached. At each step an anchor is selected, a ring size is drawn, and a candidate ring is constructed and oriented. The free-growth ring sizes are drawn with probabilities arbitrarily set as:
\begin{equation}
    P(5) = 0.15,\quad P(6) = 0.70,\quad P(7) = 0.10,\quad P(8) = 0.05\,,
\end{equation}
namely hexagon-dominated with the odd-membered rings, which are key to carrying negative curvature of disordered carbon networks \cite{martin2019topology}. These probabilities are user settings and can be extended to other ring sizes. However, the probabilities are overridden where an odd-membered ring is placed. Specifically, a candidate drawn off a 5- or 7-membered parent is redrawn either as a 6-ring or a ring of $12-n$ atoms with equal probability, until the parent has one such partner, after which its further rings are hexagons, so that a pentagon is paired with a heptagon and curvature is not accumulated by the draw alone. Neither set of weights is the ring statistics of the final structure, which merging and rejection modify during the construction and the anneal (Step 5, see Section \ref{sec:method_anneal}) modifies afterwards. The effect of other ring probabilities and growth orders on the annealed structure is not investigated in this work.

Each candidate passes two independent gates. A geometric gate ensures that (i) no atom may fall inside a pore sphere, (ii) a candidate atom closer to an existing atom than a merge threshold is fused with it rather than added, and (iii) a candidate ring whose atoms fall inside a rejection window is discarded entirely; both thresholds are multiples of the equilibrium bond length $r_0 = 1.42$ \AA{}, the first set by the merge coefficient $m$, and are exposed as model parameters. An energetic gate ensures that the candidate energy, evaluated with the model of Section \ref{sec:method_potential}, must not exceed a fixed threshold $E_\mathrm{max}$, whose value is given in the same section, with the test being applied after the ring orientation is optimized to minimize its potential energy. The seed ring, having no parent to orient against, is tested on the pore term alone. Bonds are assigned by a distance criterion at $1.2\,r_0 = 1.704$ \AA{}, applied during growth and in a final pass over the completed configuration. Growth terminates either when the target density $\rho$ is reached, which is a success condition, or when no unsaturated anchor admits any candidate, which is a geometric obstruction sensitive to the box size and pore packing rather than to the potential parameters.

\subsubsection{Interaction models during growth}\label{sec:method_potential}

Two alternative descriptions of the carbon-carbon interaction are available. The first is a
harmonic bonded model, whose energy is the sum of linear, angle and dihedral contributions, a
non-bonded 12-6 Lennard-Jones term, and a short-range 9-3 attraction acting only between
unsaturated sites:

\begin{align}
    U_\mathrm{C-C} = \sum_\mathrm{bonds}\tfrac{1}{2}k_r\left(r-r_0\right)^2 
    + \sum_\mathrm{angles}\tfrac{1}{2}k_\theta\left(\theta-\theta_0\right)^2 
    + \sum_\mathrm{dihedrals}\tfrac{1}{2}k_\varphi\sin^2\!\left(\varphi-\varphi_0\right) \nonumber \\
    + \sum_{i<j}4\varepsilon\left[\left(\frac{\sigma}{r_{ij}}\right)^{12}
    - \left(\frac{\sigma}{r_{ij}}\right)^{6}\right] 
    + \sum_{i<j}^{\mathrm{unsat}}4A_\mathrm{a}\left[\left(\frac{r_0}{r_{ij}}\right)^{9} 
    - \left(\frac{r_0}{r_{ij}}\right)^{3}\right] \, ,
\end{align}

where $r$, $\theta$ and $\varphi$ are the bond length, bond angle and dihedral angle, and $r_0$, $\theta_0 = 2\pi/3$ and $\varphi_0 = 0$ their equilibrium values, chosen so that the minimum of the bonded terms is a planar sp\textsuperscript{2} network. The fourth sum runs only over pairs separated by more than a fixed number of bonds along the connectivity graph, set to 15 here and changeable at initialization, so that it describes the interaction between distinct sheets rather than the covalent geometry already described; it is evaluated over the neighbor cells of the anchor atom, complete up to about 5 \AA{} from it, its nominal truncation being 11 \AA{}. The last sum is a growth aid biasing open fronts toward closing on one another, evaluated over the same neighbor cells, its nominal truncation being 10 \AA{}. The parameters used are $\sigma = 3.407$ \AA{} and $\varepsilon = 0.0684$ kcal mol\textsuperscript{-1} \cite{alberghini2025pore}, together with the working values $k_r = 1037.79$ kcal mol\textsuperscript{-1} \AA{}\textsuperscript{-2}, $k_\theta = 138.3724$ kcal mol\textsuperscript{-1} rad\textsuperscript{-2} and $A_\mathrm{a} = 100$ kcal mol\textsuperscript{-1}, whereas $k_\varphi$ is searched in Step 4 (see Section \ref{sec:pso}).

This model is a placement device rather than a force field for carbon, its only requirement being that a planar three-coordinated ring at $r_0$ and $\theta_0$ be its geometric minimum: the reactive potential of the anneal (Step 5) rebuilds the bond graph from the geometry and discards every parameter above.

The second available carbon-carbon interaction is the Tersoff bond-order potential \cite{tersoff1989modeling}, in which the strength of each bond is modulated by the local environment,
\begin{equation}
    U_\mathrm{C-C} = \sum_{i<j} f_C(r_{ij})\left[A_\mathrm{T}\,e^{-\lambda_1 r_{ij}} - b_{ij}\,B_\mathrm{T}\,e^{-\lambda_2 r_{ij}}\right]\,,
\end{equation}
where $f_C$ switches from one to zero between 1.8 and 2.1 \AA{}, $b_{ij}$ is the bond-order term, and $A_\mathrm{T}$, $B_\mathrm{T}$, $\lambda_1$ and $\lambda_2$ are constants of the potential. The published C-C parameterization was adopted as implemented and is listed in Supplementary Note 1. During the construction, the energy of a candidate atom is evaluated from the bond order of the atoms it binds to, which suffices as a placement criterion and does not coincide with the full Tersoff energy change. Moreover, with the bond exclusion used in this work (15 bonds, a working value like the other fixed parameters of the construction, not optimized), the anchors, the atoms within 15 bonds of them and the other atoms of the ring are left out of this energy (see Section \ref{sec:method_overview}).

Independently of the model selected, the pores act on the carbon atoms through a Lennard-Jones wall in the surface-to-surface distance,
\begin{equation}
    U_\mathrm{pore}(d) = 4A\left[\left(\frac{B}{d}\right)^{12} - \left(\frac{B}{d}\right)^{6}\right],
    \qquad d = \lVert \mathbf{r} - \mathbf{c}\rVert - R \,,
\end{equation}
for a pore of center $\mathbf{c}$ and radius $R$, summed over all pores under the minimum image convention. A 9-3 form is also available, but only the 12-6 form was used throughout. %The repulsive branch is singular as $d \to 0$, which is what allows the wall to enforce a standoff, a bounded repulsion letting carbon come to rest on the pore surface.

A candidate ring is tested on $E = U_\mathrm{C-C} + U_\mathrm{pore}$ against a single threshold $E_\mathrm{max}$. Since the two terms are summed, the threshold and the well depth $A$ of the pore wall carry the energy unit of the carbon-carbon model they accompany, namely kcal mol\textsuperscript{-1} with the harmonic model and eV with Tersoff, whereas the length $B$ is in \AA{} in both cases. The threshold is a user setting, here $E_\mathrm{max} = 10^{3}$ in either unit. The pore wall acts in this full form during the construction only, its conversion and truncation for the anneal being given in Section \ref{sec:method_anneal}. % This is the only construction quantity which survives into the anneal, and the conversion it requires is given in Section 5. The complete parameter list, with those held fixed and those searched, is given in Supplementary Note 1.

\subsection{Structure characterization}\label{sec:post-process}

Every structure, whether constructed or read from a trajectory, is characterized by the same routines through the same code path, so that a target and a candidate are never compared across differing conventions. The fingerprint comprises the density, the pore size distribution, the porosity, the accessible surface area and volumes, the radial distribution function, the coordination and ring statistics, and the pore connectivity.

The void measures are built on a single nearest-atom distance field, computed once per structure on a voxel grid of constant spacing $dx$, which sets the resolution of every quantity below. Two conventions are shared across the algorithms. The geometric measures are periodic, the nearest-atom search including the periodic images of every atom. The accessibility measures are deliberately not: a periodic cell has no exterior, so the box faces are taken as the interface through which an adsorbate enters the sample, and the accessible void is the region which a flood fill seeded at the faces reaches. Void space is classified by two scalars, a carbon radius $r_C$ and a probe radius $r_N$, a voxel counting as void when its distance to the nearest atom center exceeds $r_C + r_N$. Throughout the generation and optimization Steps $dx = 0.5$ \AA{}, $r_C = 1.71$ \AA{} and $r_N = 1.55$ \AA{} were considered, the latter being the van der Waals radius of nitrogen \cite{bondi1964van}, so that the reported void is the one accessible to a nitrogen probe. The post-processing validation (see Section \ref{sec:validation} and Supporting Information) uses the different set matching PoreBlazer's \cite{sarkisov2020materials}; the two are never mixed and both are listed in Supplementary Note 9.

\textbf{Pore size distribution.} The method is based on a Monte-Carlo probe insertion on the voxel grid, the size reported for each insertion point being the diameter of the largest probe-accessible sphere containing it, the accumulated cumulative distribution being then differentiated. This follows the probe-insertion definition of Gelb and Gubbins \cite{gelb1999pore} in the form given by Bhattacharya and Gubbins \cite{bhattacharya2006fast}, and the implementation of PoreBlazer \cite{sarkisov2020materials}, on which this routine is modeled.

\textbf{Porosity and volumes.} The porosity is the periodic void fraction at threshold $r_C + r_N$, with no accessibility requirement, a channel which the objective can score but which was not scored in the campaigns reported here. The geometric pore volume $V_G$ is the void fraction at zero probe radius, expressed per unit mass; it is reported as an unscored property and is also one of the quantities exposed by PoreBlazer. Two further volumes come from the non-periodic accessibility analysis: the probe-centre accessible volume $V_\mathrm{centre}$, namely the voxels beyond $r_C + r_N$ which the flood fill connects to a box face, and the probe-occupiable volume $V_\mathrm{PO}$, obtained by dilating $V_\mathrm{centre}$ by the probe radius, which is the quantity adsorption experiments report as pore volume.

\textbf{Accessible surface area.} A triangulated isosurface of the distance field at level $r_C + r_N$, extracted by marching cubes \cite{lorensen1998marching} in the implementation of \verb+scikit-image+ \cite{van2014scikit}, with the free but pore-enclosed voxels excluded so that only the network-accessible surface is measured. Counting exposed voxel faces instead would converge to the Minkowski and not to the Euclidean area (see Supplementary Note 9). The dilation defining $V_\mathrm{PO}$ uses the Euclidean distance transform of \verb+scipy.ndimage+ \cite{virtanen2020scipy}, namely a spherical and not a cubic probe.

\textbf{Coordination and rings.} Two atoms are bonded if their distance is below $1.2\,r_0 = 1.704$ \AA{}, a user-defined threshold, the same rule rebuilding the bond graph after generation and when an annealed trajectory is loaded. On this graph the primitive rings, namely those containing no shortcut path among their members \cite{franzblau1991computation}, are enumerated by a bounded search \cite{yuan2002efficient} up to 12 atoms. Coordination fractions are not comparable across bond cutoffs, so every fraction reported here is at 1.704 \AA{}; a literature value obtained at another cutoff, for instance the 1.85 \AA{} of de Tom\'as et al. \cite{de2017structural}, is compared by re-processing the structures at that cutoff.

\textbf{Connectivity.} A Voronoi tessellation of the atomic positions \cite{barber1996quickhull} supplies the candidate pores, namely its vertices farther than $r_C + r_N$ from every atom, and the candidate links between them, namely its edges, a link being kept only if the probe fits along it. $k_\mathrm{avg}$ is the number of links per pore, each link crossing a periodic face being counted once.

\subsection{Structure--target distance: costing a proxy structure}\label{sec:costing}

A candidate is scored against the target properties by a weighted sum of per-channel discrepancies, the available channels being the density, the accessible surface area, the accessible pore volume, the void fraction, the pore size distribution together with its fraction below a stated diameter, and the radial distribution function:
\begin{equation}
    \mathrm{Cost} = \sum_{j} w_j\, c_j \, ,
\end{equation}
with $w_j = 1$ throughout this work unless stated otherwise. For a scalar channel the discrepancy is the relative deviation from the target, which is what makes channels carrying different units commensurable,
\begin{equation}
    c_j \;=\; \frac{\left|x_j - x_j^{*}\right|}{x_j^{*}} \, ,
\end{equation}
where $x_j$ is the measured value and $x_j^{*}$ the target. Two channels are measured on the whole cell and reported per unit mass, as in experimental and numerical practice, so they are converted first,
\begin{equation}
    \mathrm{ASA} = \frac{10^{4}\,S}{\rho\,L^{3}} ,
\qquad V_\mathrm{PO} = \frac{V}{\rho\,L^{3}} \, ,
\end{equation}
with ASA in m\textsuperscript{2} g\textsuperscript{-1} and $V_\mathrm{PO}$ in cm\textsuperscript{3} g\textsuperscript{-1}, $S$ the accessible surface of the cell in \AA{}\textsuperscript{2}, $V$ its probe-occupiable volume in \AA{}\textsuperscript{3}, $\rho$ the achieved density in g cm\textsuperscript{-3} and $L$ the box edge in \AA{}.

The pore size distribution contributes an earth-mover distance \cite{rubner2000earth}. Each differential distribution is converted to a cumulative distribution over the pore radius and normalized to unit mass, and the distance is the area between the two curves,
\begin{equation}
    W_1 \;=\; \int_0^{\infty} \left| F(r) - F^{*}(r) \right| \,\mathrm{d}r \, ,
\end{equation}
divided by the density-weighted mean radius of the target so that the channel is dimensionless. $W_1$ penalizes a distribution with the right mean but the wrong shape, which a comparison based on moments alone would accept. Because it compares cumulative curves, a deficit at small radii and an excess at large radii partially cancel, so a second channel is added which does not cancel, namely the fraction of pore volume below a stated diameter $d^{*}$,
\begin{equation}
    f(d^{*}) = \int_0^{d^{*}} \frac{\mathrm{d}V}{\mathrm{d}D}\,\mathrm{d}D 
    \left( \int_0^{\infty} \frac{\mathrm{d}V}{\mathrm{d}D}\,\mathrm{d}D\right)^{-1} ,
\qquad c_\mathrm{psd,low} = \left| f(d^{*}) - f^{*}(d^{*}) \right| \, ,
\end{equation}
with $d^{*}$ accurately chosen to isolate the smallest mode of the target and reported with each campaign.

The radial distribution function is sampled locally (see Section \ref{sec:method_overview}). $N_s = 20$ points are drawn uniformly at random in the cell, and the atoms within 7 \AA{} of each point are taken as the central atoms of one local $g(r)$. A point with no atom within this distance, which falls inside a pore, gives no curve and is discarded, so that the channel uses at most $N_s$ curves. The neighbors of the central atoms are taken from the whole periodic structure, so the local curves are not truncated at the sphere boundary, and the normalization retains the global number density $\rho_\mathrm{N} = N/L^{3}$ of the $N$ atoms of the cell,
\begin{equation}
    g_\mathrm{loc}(r) = \frac{n(r)}{\rho_\mathrm{N}\, N_\mathrm{c}\, V_\mathrm{shell}(r)} \, ,
    \qquad V_\mathrm{shell}(r) = \frac{4\pi}{3}\left[\left(r+\Delta r\right)^{3} - r^{3}\right] ,
\end{equation}
where $n(r)$ is the number of pairs formed by a central atom and any atom at a distance between $r$ and $r + \Delta r$, $\Delta r$ the bin width and $N_\mathrm{c}$ the number of central atoms. Since $\rho_\mathrm{N}$ is that of the whole cell, an under-dense pocket plateaus below unity rather than being rescaled to one. Each
local curve is compared with the reference by a radius-averaged absolute difference over the range both curves cover, $r_\mathrm{min}$ being their first common bin,
\begin{equation}
    D\left(g,g^{*}\right) = \frac{1}{r_\mathrm{max}-r_\mathrm{min}}
\int_{r_\mathrm{min}}^{r_\mathrm{max}} \left| g(r) - g^{*}(r) \right| \mathrm{d}r ,
\qquad r_\mathrm{max} = 7\ \text{\AA} \, ,
\end{equation}
and the channel is the mean of the worst quartile of these distances. %,
%
%\begin{equation}
%    c_\mathrm{rdf} = \frac{1}{k}\sum_{i=1}^{k} D_{(N_s-i+1)} , \qquad k = \mathrm{round}(0.25\,N_s) \, .
%\end{equation}
%
Finally, before comparison with the reference, the candidate RDF curve is convolved with a row-normalized Gaussian of width $\sigma = 0.085$ \AA{}, which compensates the absence of thermal motion in a proxy structure. The width is not a fitting parameter, being the root-mean-square spread of the first neighbor distance of amorphous carbon at room temperature \cite{li1990radial}. The smearing is applied at comparison time only.

Each channel carries a hard tolerance: if any deviates from its target by more than $\delta_\mathrm{max} = 0.9$, the cost is returned as non-finite rather than as a large number, as is the cost of a failed generation. A large but finite penalty would leave a gradient pointing into the inadmissible region, whereas a non-finite cost removes the candidate from the memory of the swarm. Each candidate is evaluated with seven independent random seeds and scored by the mean over the replicates that completed, at least five being required for it to take part in the swarm update, so that a single anomalous replicate can neither reject a good parameter set nor promote a bad one; both numbers can be set by the user.

\subsection{Surrogate-assisted particle swarm optimizer (PSO)}\label{sec:pso}

The generation parameters producing structures closest to the prescribed targets are recovered by particle swarm optimization \cite{kennedy1995particle}, augmented with a response-surface surrogate. 
Each particle is a point in the bounded parameter space and each evaluation is a complete construction (Step 1, see Section \ref{sec:structure_generation}) followed by a complete characterization (Step 2, see Section \ref{sec:post-process}), scored by the objective of Step 3 (see Section \ref{sec:costing}). Writing $\mathbf{x}_i^{t}$ and $\mathbf{v}_i^{t}$ for the position and velocity of particle $i$ at iteration $t$, $\mathbf{p}_i$ for its own best position and $\mathbf{g}$ for the swarm best, the update is:
\begin{equation}
    \mathbf{v}_i^{t+1} = \underbrace{w\,\mathbf{v}_i^{t}}_{\text{inertia}} + \underbrace{c_1\,\mathbf{r}_1 \odot \left(\mathbf{p}_i - \mathbf{x}_i^{t}\right)}_{\text{cognitive}} + \underbrace{c_2\,\mathbf{r}_2 \odot \left(\mathbf{g} - \mathbf{x}_i^{t}\right)}_{\text{social}} ,
    \qquad \mathbf{x}_i^{t+1} = \mathbf{x}_i^{t} + \mathbf{v}_i^{t+1} \, ,
\end{equation}
where $\odot$ is the element-wise product and $\mathbf{r}_1$, $\mathbf{r}_2$ are uniform random vectors drawn at every iteration and component, which vary the strength of each pull so that the particles do not move along straight lines and keep exploring. The inertia weight $w$ sets the balance between exploration and convergence \cite{shi1998modified}, while $c_1$ and $c_2$ weight the attraction toward the particle's own best and the swarm best. Each velocity component is capped at a fraction $v_\mathrm{max}$ of its parameter range and a particle reaching a bound is clipped to it with that velocity component reset to zero. The values used throughout are $w = 0.7$, $c_1 = c_2 = 1.5$, $v_\mathrm{max} = 0.2$, with 16 particles over 25 iterations.

Beyond a warm-up threshold, a forward model $\hat{f}$ of the cost is refitted at each iteration on the whole accumulated history: a degree-two polynomial expansion of the parameters, each resulting feature being standardized, followed by an elastic-net regression \cite{zou2005regularization},
\begin{equation}
    \boldsymbol{\beta} = \arg\min_{\boldsymbol{\beta}} \left\{
\frac{1}{2M}\sum_{k=1}^{M} u_k \left( C_k - \hat{f}_{\boldsymbol{\beta}}(\mathbf{x}_k) \right)^2
+ \alpha\,\ell_1 \lVert\boldsymbol{\beta}\rVert_1
+ \frac{\alpha\left(1-\ell_1\right)}{2}\lVert\boldsymbol{\beta}\rVert_2^{2} \right\} \, ,
\end{equation}
where $M$ is the number of evaluations accumulated, $C_k$ the mean cost of candidate $k$, $\alpha$ the overall penalty strength and $\ell_1$ its split between the two norms. The sample weights $u_k = 1/(\sigma^{2}_{r,k} + \epsilon)$ are the inverse replicate variances, so that a noisily measured point informs the fit less, with $\epsilon$ a floor. The values used are $\alpha = 10^{-3}$, $\ell_1 = 0.5$, $\epsilon = 10^{-3}$, a warm-up of 64 successful evaluations. Minimizing $\hat{f}$ over the bounds from 15 random restarts with a quasi-Newton method \cite{virtanen2020scipy} yields one additional candidate per iteration, evaluated alongside the swarm only through $\mathbf{g}$ and only if strictly better than the current global best.

\subsubsection{PSO parameters identifiability} 

The whole evaluation history, including the candidates rejected as inadmissible, is read back as a designed experiment sampling the cost surface, to estimate the practical identifiability of each parameter, namely whether the data collected constrained it or left it free \cite{raue2009structural}; the description in terms of stiff and sloppy directions follows Refs. \cite{transtrum2015perspective,gutenkunst2007universally}.

A full second-order polynomial in the standardized parameters is fitted by ordinary least squares to the costs of the admissible candidates of iterations 1 onward,
\begin{equation}
    \hat{c}(\tilde{\mathbf{x}}) \;=\; c_0 + \mathbf{b}^{\mathsf{T}}\tilde{\mathbf{x}}
+ \tfrac{1}{2}\,\tilde{\mathbf{x}}^{\mathsf{T}}\mathbf{H}\,\tilde{\mathbf{x}} \, ,
\end{equation}
where $\tilde{x}_j = (x_j - \mu_j)/s_j$ is the $j$-th of the $D$ free parameters, standardized by its mean $\mu_j$ and standard deviation $s_j$ over the fitted candidates, $\mathbf{b}$ collects the linear coefficients and the symmetric matrix $\mathbf{H}$ of the quadratic and cross terms is the Hessian of the fitted surface. The random initial swarm is excluded from the fit, since its costs, several times those of the converged swarm and far from the region the swarm later explores, would otherwise set the fitted curvature by their leverage alone. Three quantities are read from this fit.

The first is its quality, measured by the coefficient of determination $R^2$ of the fit. Each candidate is evaluated in up to seven independent replicates and scored by their mean cost, so part of the cost variance is replicate noise, whose share is:
\begin{equation}
    \eta = \frac{\big\langle\, \sigma^{2}_{r,k} / n_k \,\big\rangle_k}{\mathrm{Var}_k\!\left(c_k\right)} \, ,
\end{equation}
where $\sigma^{2}_{r,k}$ and $n_k$ are the replicate variance and the number of surviving replicates of candidate $k$, $c_k$ its mean cost and $\langle\cdot\rangle_k$ the average over the candidates. No function of the parameters can explain this share, so when $1 - R^{2}$ is much larger than $\eta$ the variance the fit misses is structure of the cost rather than noise. Both quantities are conditional on the candidates actually sampled, and no standard error is taken from the regression: the uncertainty of the curvatures is estimated by resampling, as described below.

The second is the curvature of the cost. The eigendecomposition:
\begin{equation}
    \mathbf{H} = \mathbf{V}\boldsymbol{\Lambda}\mathbf{V}^{\mathsf{T}}, \qquad \boldsymbol{\Lambda} = \mathrm{diag}(\lambda_1,\dots,\lambda_D) \, ,
\end{equation}
%
%computed with a symmetric solver (\verb+numpy.linalg.eigh+ \cite{harris2020array}), 
returns the principal curvatures $\lambda_i$ of the cost and the parameter combinations $\mathbf{v}_i$ along which they act. Since the fit is global over the sampled region, the eigenvalues describe the sampled surface rather than a strict neighborhood of the optimum. Its uncertainty is estimated by resampling whole particle trajectories, since the successive positions of a particle are not independent: the $P$ particles are drawn with replacement, the fit, standardization included, is repeated on the resampled history, and the eigenvalues are ranked from the largest. Over 1000 resamples this gives a 95 \% percentile interval for each ranked curvature and the fraction of resamples in which that direction is dominated by the same parameter as in the full fit, on which the robustness of an attribution is judged. Since the particles are few and coupled through the swarm best, these intervals underestimate the true uncertainty.

The third is the share of the cost variance each parameter accounts for, given by the drop-one partial coefficient of determination:
\begin{equation}
    R^{2}_{\mathrm{part},j} \;=\; R^{2}_{\mathrm{full}} - R^{2}_{-j} \, ,
\end{equation}
being $R^{2}_{-j}$ the coefficient of determination of the same quadratic refitted without the $j$-th parameter, that is without its linear, square and cross terms, so that $R^{2}_{\mathrm{part},j}$ is the fraction of the total cost variance which parameter $j$ explains uniquely. % These contributions do not need to sum to $R^{2}_{\mathrm{full}}$: variance explained jointly by correlated parameters is attributed to none of them, whereas variance carried by the interaction term of two parameters is removed with either of them and counted twice. 

A parameter is classified as stiff when it explains at least 5 \% of the cost variance uniquely ($R^{2}_{\mathrm{part},j} \geq 0.05$) and has the largest component, in absolute value, of an eigenvector of positive curvature, the attribution being robust when that eigenvector keeps the same dominant parameter in at least 85 \% of the resamples; both thresholds are conventions of this work rather than statistical tests. Any other parameter is either sloppy or boundary-limited, depending on its recovered value: in the interior of the search range the data is indifferent to it (sloppy), whereas at a bound where the candidates of the history accumulate its flat or negative curvature reflects a gradient pushing into the bound (boundary-limited). A run is taken as identified when at least one parameter is stiff with a robust attribution; a search in which none is has stopped on a region of the cost which the data do not constrain, rather than at an optimum.

\subsection{Anneal of constructed structures}\label{sec:method_anneal}

Constructed structures are relaxed with LAMMPS \cite{thompson2022lammps}, the carbon-carbon interaction being described by the EDIP parameterization of Marks \cite{marks2000generalizing}, accessed through the OpenKIM portable model \verb+EDIP_LAMMPS_Marks_2000_C__MO_374144505645_000+ \cite{tadmor2011potential}.
The library writes each structure directly as LAMMPS input: a data file of the atoms, and one spherical region per pore acted on by a \verb+fix wall/region+ command, replicated to every periodic image whose influence sphere still intersects the cell. The wall is the 12-6 pore potential of the construction with the same $A$ and $B$, its depth being converted to eV where the harmonic model sets it in kcal mol\textsuperscript{-1}. Each wall is truncated at its own minimum, $d = 2^{1/6}B$, so that only its repulsive branch is retained, in the manner of the Weeks-Chandler-Andersen (WCA) potential \cite{weeks1971role}.

The proxy structure is integrated with a timestep of 0.1 fs in the four stages of Table \ref{tab:method_anneal}. A conjugate-gradient minimization preceding the dynamics did not progress beyond its first step, owing to a defect in the energy reported by the region walls in the LAMMPS release used. For the same reason, the potential energy is reported for the EDIP potential alone, and the wall energy is recomputed from the coordinates of the initial and final structures. The relaxed start tested in Section \ref{sec:anneal} uses the FIRE algorithm \cite{bitzek2006structural,guenole2020assessment} (Supplementary Note 7). 

\begin{table}[h]
\centering
\caption{LAMMPS annealing procedure}\label{tab:method_anneal}
\begin{tabular}{@{}lll@{}}
\toprule
    stage & ensemble / thermostat & duration \\
\midrule
     Burn-in &  NVE + Langevin \cite{schneider1978molecular}, 300 K $\to$ \textbf{T}, damping 0.01 ps & 1 ps \\
     Hold & Nos\'e-Hoover NVT, \textbf{T}, damping 0.01 ps & 20 ps \\
     Quench & Nos\'e-Hoover NVT, \textbf{T} $\to$ 300 K & 17.6 ps \\
     Cold hold & Nos\'e-Hoover NVT, 300 K & 0.5 ps \\
\end{tabular}
\end{table}
The quench is run at a fixed 80 steps per kelvin, namely a rate of 1.25$\cdot$10\textsuperscript{14} K s\textsuperscript{-1}, a working choice of this work, so its duration scales with the setpoint and is 17.6 ps for \textbf{T} = 2500 K; the total anneal is then 39.1 ps. The choice of the thermostats and of the walls is motivated in Section \ref{sec:anneal}, and the tests which set the protocol are reported in Supplementary Note 7.
The volume is held fixed throughout. % Density is a construction constraint in this workflow, and an isobaric ensemble would relax the cell and collapse the pores by volume change, confounding precisely the quantity under test. 
Coordination is read from the cold final frame, since a hot frame lowers the neighbor count. A structural quantity is taken as converged when its change over the last 5 ps of the hold is below 10 \% of its whole change between the proxy structure and the end of the hold, its convergence time $t_{90}$ being the time at which it completes 90 \% of that change. To separate the effect of the protocol from that of the construction, the 100 \AA{} reference structure of Section \ref{sec:qmd_protocol} was annealed with the same protocol, without walls.

\subsection{Validation of the post-processing}\label{sec:validation}

The characterization routines of Section \ref{sec:post-process}, except the pore connectivity, which has no external counterpart, were validated channel by channel against independent implementations, at matched conventions (bond cutoff, probe radii, bin widths and ring definition): the pore size distribution and the accessible surface and volumes against PoreBlazer v4.0 \cite{sarkisov2020materials}, the radial distribution function against OVITO \cite{stukowski2010visualization}, the coordination against MDAnalysis \cite{michaud2011mdanalysis} and the ring statistics against R.I.N.G.S. v1.3.5 \cite{le2010ring}. Two frames were used, the 100 \AA{} reference of Section \ref{sec:qmd_protocol} and a 50 \AA{} frame from a separate quench protocol (Supplementary Note 9), whose thermal history does not enter a comparison of two implementations reading the same coordinates. The coordination and the RDF agree to machine precision, the ring statistics within 0.01, the PSD within 0.1 \AA{} and the pore volumes within 2.5 \%, whereas the accessible surface area is 15 to 25 \% above the reference value depending on the structure. The two implementations reconstruct the same surface by different means, a triangulated level set of a sampled distance field here and a stochastic per-atom sampling in PoreBlazer, and their deviation widens rather than closes as the grid is refined, which identifies a difference of algorithm rather than a discretization error; it is therefore reported rather than corrected. The search of Step 4 is unaffected, since target and candidate surface areas are computed by the same implementation at the same parameters. The comparison design, the matched parameters, the per-channel numbers and the grid-convergence study are given in Supplementary Note 9.

\subsection{Generation of reference QMD structures}\label{sec:qmd_protocol}

Reference structures were produced by quenched molecular dynamics from a disordered initial state, which is the standard route to amorphous carbon \cite{palmer2010modeling}. The initial configurations were random carbon packings at $\rho = 0.95$ g cm\textsuperscript{-3} with a minimum interatomic separation of 1.5 \AA{}, in cubic periodic boxes of 100, 150, 200 and 250 \AA{} containing 47,640, 160,784, 381,117 and 744,369 atoms, namely a 15.6-fold range in atom count.

Simulations were performed with LAMMPS \cite{thompson2022lammps}, using the environment-dependent interatomic potential (EDIP) for carbon \cite{marks2000generalizing} through its OpenKIM portable model \cite{tadmor2011potential}, with a timestep of 0.1 fs. The thermal protocol was identical for every box and was carried out in the canonical ensemble at a damping time of 0.1 ps: 20 ps at 2500 K, a 320 ps linear ramp from 2500 K to 300 K at 6.9 K ps\textsuperscript{-1}, and 30 ps at 300 K, for a total of 370 ps. All four runs completed the full 3.7 million steps and their final configurations were written at 298 to 300 K; every structural analysis reported here was performed on those cold frames.

The protocol is size-converged in both energy and local structure. The final potential energy lies between -6.728 and -6.733 eV/atom across the structures generated, and the three-coordinated fraction, measured at a 1.704 \AA{} cutoff, is 80.9, 80.9, 80.6 and 80.5 \% at 100, 150, 200 and 250 \AA{}. %, a spread of 0.4 percentage points across a 15.6-fold range in atom count. At 100 \AA{} the full breakdown is 12.2 \% two-, 80.9 \% three- and 6.9\% four-coordinated, with a nearest-neighbor distance of 1.417 \AA{}.

\subsection*{Statistics and reproducibility}
Each candidate of the inverse design is evaluated with seven independent random seeds and scored by the mean cost over the replicates that complete (Section \ref{sec:costing}). The anneal is applied to three replicate structures of each of the four runs, and the values reported after the anneal are means over these replicates, the spread being given as the standard deviation or as the range over replicates or runs, as stated in each caption and table. The intervals of the identifiability analysis are 95 \% percentile intervals over 1000 bootstrap resamples of whole particle trajectories (Section \ref{sec:pso}). No hypothesis test is performed, the thresholds used to classify the parameters being conventions of this work. The data from which the reported values are computed are listed under Data availability.

\subsection*{Use of Large Language Models}
A large language model was used to assist with three tasks. In writing, it edited the language of the manuscript and Supplementary Information, drafting sections from provided outlines, which the authors then revised for content and tone. In numerical analysis, the model wrote notebooks to compute observables from raw data and verify the numbers reported in the manuscript against the simulation output. In the formal analysis, it was also used to search for and summarise literature references, as well as to discuss, verify and improve analysis options. The authors chose, checked and interpreted the adopted options and the results obtained. Every reference was checked by the authors against the original source. The authors take full responsibility for the content of this article.

\subsection*{Code availability}
Steps 1--4 of the methodology are implemented in \verb+nanocarbon+, a Python library developed for this purpose and released under the Creative Commons Attribution-NonCommercial-ShareAlike 4.0 International license (CC BY-NC-SA 4.0) at \url{https://github.com/Gemmate-Technologies/nanocarbon} and on the Python Package Index. The deposited scripts run on its release 1.0.0, which remains available from both. The anneal was run with LAMMPS \cite{thompson2022lammps}, whose input deck is deposited with the data, and the external tools used to validate the post-processing are publicly available (Section \ref{sec:validation}).

\subsection*{Data availability}
The data supporting the findings of this study are deposited at \url{https://doi.org/10.5281/zenodo.23315448} in five archives: (i) the inverse-design campaigns, namely the configuration files, the complete optimization histories, the structures generated from the recovered parameter vectors and the identifiability analysis; (ii) the anneal of these structures and of the QMD reference, namely the LAMMPS input deck and submission scripts, the logs, the structures before and after the anneal, the per-frame structural analysis and the tests which set the protocol; (iii) the annealing trajectories; (iv) the input decks, logs, initial and final configurations of the reference structures obtained via QMD; (v) the scripts comparing the post-processing with the external tools, together with the outputs of these tools. These files, together with the notebooks of the analysis, contain the data from which the values reported in this article are computed. The data are released under the Creative Commons Attribution 4.0 International license (CC BY 4.0), and the scripts and the notebooks under CC BY-NC-SA 4.0.

\subsection*{Acknowledgements}
This research was performed within the HEDAsupercap project, funded by the European Union's Horizon Europe Research and Innovation Actions under grant agreement no. 101092189. The funder played no role in study design, data collection, analysis and interpretation of data, or the writing of this manuscript. The authors thank Dr. Fabrizio Silveri for the discussions in the early stage of the project.

\subsection*{Author contributions}
M.Al. conceived the study, developed the nanocarbon library, performed the inverse-design campaigns, the anneals and the analysis, and wrote the manuscript. G.P. contributed to the development of the nanocarbon library and revised the manuscript. M.Am. performed the quenched molecular dynamics simulations of the reference structures and contributed to the definition of the annealing protocol. A.T. performed the funding acquisition and managed the research project. A.T. and M.S. supervised the work. All authors read and approved the final manuscript.

\subsection*{Competing interests}
The authors declare no financial or non-financial competing interests.

\bibliographystyle{unsrt}
\bibliography{references}

\end{document}